\documentclass[fleqn,usenatbib]{mnras}

\usepackage{mathptmx}

\usepackage[T1]{fontenc}

\DeclareRobustCommand{\VAN}[3]{#2}
\let\VANthebibliography\thebibliography
\def\thebibliography{\DeclareRobustCommand{\VAN}[3]{##3}\VANthebibliography}

\usepackage{graphicx}	
\usepackage{amsmath}	
\usepackage{txfonts}   
\usepackage{multirow}  
\usepackage{tablefootnote} 
\usepackage{threeparttable}  
\usepackage{orcidlink}
\usepackage{multicol}
\newcommand{\rev}[1]{{\textcolor{black}{#1}}}

\title[Beyond the FMR]{Beyond the Fundamental Metallicity Relation: galaxy sizes encode the link between inflow and metallicity} 

\author[Boardman et al.]{
N.~F.~Boardman$^{1}$\thanks{E-mail: nfb@st-andrews.ac.uk}\,\orcidlink{0000-0002-9119-292X},
V.~Wild$^{1}$\,\orcidlink{0000-0002-8956-7024}, 
D. ~Scholte$^{2}$, 
K. ~Wang$^{3,4}$\,\orcidlink{0000-0002-3775-0484},
N. ~Vale Asari$^{5}$,
A. ~Saintonge$^{6,7}$\\
$^{1}$School of Physics and Astronomy, University of St Andrews, North Haugh, St Andrews KY16 9SS, UK\\
$^{2}$Institute for Astronomy, University of Edinburgh, Royal Observatory, Blackford Hill, Edinburgh EH9 3HJ, UK\\
$^{3}$Institute for Computational Cosmology, Department of Physics, Durham University, South Road, Durham, DH1 3LE, UK\\
$^{4}$Centre for Extragalactic Astronomy, Department of Physics, Durham University, South Road, Durham DH1 3LE, UK\\
$^{5}$Departamento de F\'{\i}sica--CFM, Universidade Federal de Santa Catarina, C.P.\ 5064, 88035-972, Florian\'opolis, SC, Brazil \\
$^{6}$Department of Physics \&\ Astronomy, University College London, London, UK\\
$^{7}$Max Planck Institute for Radio Astronomy, Auf dem Hugel 69, 53121 Bonn, Germany
}

\date{Accepted 2026 June 29. Received 2026 June 8; in original form 2026 April 7}
\pubyear{2026}
\begin{document} 
\label{firstpage}
\pagerange{\pageref{firstpage}--\pageref{lastpage}}
\maketitle

\begin{abstract}
Gas-phase chemical abundances are key observable consequences of galaxy evolution, being intrinsically tied to galaxy formation histories. Gas metallicity rises with increasing stellar mass ($\mathrm{M_*}$), forming the well-known mass-metallicity relation (MZR). MZR residuals have separately been shown to anti-correlate with star-formation rate (the ``fundamental'' metallicity relation), with gas mass and with optical size, but no single analysis has considered all trends together. We thus perform a combined analysis of all three trends, utilizing optical MaNGA integral field spectroscopy, HI-MaNGA gas masses, and MaNGA DynPop dynamical masses. We estimate inner gas masses for $\sim$1500 star-forming galaxies, finding this to be the most important parameter after $\mathrm{M_*}$ in predicting gas metallicities. We obtain equivalent results for stellar metallicities and gaseous N/O, suggesting that current inner gas masses are intrinsically linked to long-term chemical evolution histories. We show that more compact galaxies have lower dynamical masses, challenging suggestions that deeper gravitational potentials confer higher metallicities. We find a strong correlation between inner gas mass and galaxy size, meaning that short term inflow fluctuations cannot be responsible for the MZR residuals. With chemical evolution models, we show that our results can instead be explained by differences in long-term inflow histories. The earlier inflow histories of compact galaxies lead to lower gas masses and more rapidly declining gas reservoirs at late times, leading to higher metallicities. At fixed stellar mass, galaxy size therefore encodes the link between halo assembly histories, long-term gas inflow histories, current gas reservoirs and metallicity.

\end{abstract}


\section{Introduction}\label{intro}


Galaxy chemical abundances are key observable consequences of galaxy evolution. Star-formation leads to the production of fresh metals and subsequent injection of those metals into the interstellar medium (ISM); gaseous inflows provide fuel for star-formation and dilute the metallicity of star-forming gas, while outflows --- driven by star-forming feedback or else by feedback from AGN --- reduce star-formation and remove enriched material \citep[e.g.][]{schmidt1963,finlator2008,forbes2014,peng2014,bb2018}. Galaxy chemical abundances thus provide a window into how galaxies form and evolve and are directly sensitive to star-formation histories (SFHs) and gas accretion histories, making them a key benchmark for models and simulations \citep[for a review see][]{maiolino2019}.

Of all chemical abundance indicators, the gaseous oxygen abundance --- hereafter gas metallicity, 12 + $\mathrm{\log(O/H)}$ --- is the most commonly studied. This can be determined within star-forming galaxies with relatively little difficulty, using bright emission lines in their optical spectra \citep[e.g.][and references therein]{curti2020}. Oxygen is produced mostly in massive ($\mathrm{>8\mathrm{M_\odot}}$) stars and then released on short ($\sim$10 Myr) timescales by core-collapse supernovae \citep[e.g.][]{timmes1995,johnson2019,kobayashi2020}. Gas metallicities are therefore expected to be very sensitive to star-formation, as well as being sensitive to the metallicity of gaseous inflow on short timescales \citep[e.g.][]{hwang2019,luo2021,lazarus2026}.

A mass-metallicity relation (MZR) is observed across star-forming galaxy populations: galaxies with higher stellar masses ($M_*$) possess metal-richer gas on average, with metallicity approaching a plateau at the high-mass end. This is seen both in the local Universe \citep[e.g.][]{lequeux1979,tremonti2004,andrews2013,scholte2024,scholte2026} and at higher redshifts \citep[e.g.][]{erb2006,maiolino2008,zahid2011,jones2020,pallottini2025}, with higher-redshift galaxies being offset to progressively lower metallicities at fixed $M_*$. Researchers have proposed various explanations for the MZR, including decreased outflow effectiveness with increasing mass \citep[e.g.][]{tremonti2004,tortora2022} or else increased gas reservoir consumption at high stellar masses \citep[e.g.][]{spitoni2020,scholte2024}. However, the scatter in the MZR is significantly larger than can be expected from observational errors alone, suggesting that stellar mass is not the only driver of gaseous metallicities \citep[e.g.][]{hoopes2007}. We can learn much more, therefore, by considering what other galaxy properties correlate with metallicity.

Following \citet{tremonti2004}, it was quickly established that both star-formation rate (SFR) and galaxy size correlate with galaxies' MZR residuals \citep{ellison2008}: at a given stellar mass, higher metallicities are associated with lower SFRs and smaller sizes. The SFR connection especially captured the community's attention, and it is today referred to as the fundamental metallicity relation \citep[FMR;][]{ll2010,mannucci2010}. The FMR has been studied at length within the nearby Universe \citep[e.g.][]{andrews2013,ll2013,sanchez2017a,curti2020} and is found to hold out to $z \simeq 2.5$ \citep[e.g.][]{sanders2018,cresci2019}, leading to suggestions of it being truly fundamental. The FMR seems to shift to lower metallicities at $\mathrm{z \gtrsim 2.5}$ \citep{troncoso2014,curti2023,nakajima2023,heintz2023,curti2024,scholte2025,stanton2025}, with high-z analogs also offset to lower metallicities from the FMR \citep{yang2017,yang2017a,stanton2025}; much of this however relies on extrapolations in $\mathrm{M_*}$--SFR space from low-redshift relations \citep[e.g.][]{laseter2025}, with the size of the offset also depending on the adopted form of the FMR \citep[e.g.][]{scholte2025}. It is also important to note that the FMR is subtle: metallicities anti-correlate with SFR only mildly at fixed $\mathrm{M_*}$ \citep{sanchez2013,bb2017,salim2014,sanchez2017a,sanchez2019a,ma2024a,boardman2025,kc2025}, with \citet{laseter2025} reporting no detection of the FMR for galaxies below $\mathrm{\sim
10^9\ M_\odot}$; this leaves most MZR scatter unaccounted for, meaning that the FMR is unlikely to be fundamental in a physical sense. However, the three parameters of the FMR --- $\mathrm{M_*}$, SFR and gas metallicity --- are relatively easy to observe even at high redshifts, making the FMR a common point of comparison between models and simulations \citep[e.g.][]{dave2017,derossi2017,torrey2019,ma2024a,garcia2025}. 
 
The FMR is most commonly interpreted in terms of gas reservoirs, with variations in gas flows leading to variations in SFR and opposite variations in metallicity. The FMR has been reproduced from such a framework in analytical models \citep[e.g.][]{lilly2013,forbes2014,zahid2014,wang2021}, semi-analytical models \citep[e.g.][]{yates2012,delucia2020} and hydrodynamical simulations \citep[e.g.][]{derossi2017,torrey2018,vanloon2021}. In agreement with this view, MZR residuals have been found to show significant anti-correlations with atomic and/or molecular gas masses \citep{hughes2013,bothwell2013,bothwell2016,bothwell2016a,brown2018,chen2022a,scholte2023,scholte2024}. In the case of HI gas, the MZR's scatter becomes even smaller when using HI masses estimated within galaxies' optical radii \citep{chen2022a} instead of global HI masses. All of these points support a view in which metallicity variations at fixed stellar mass are tied to galaxy gas reservoirs, with the FMR emerging as a consequence.

In isolation, the FMR can be explained either in terms of rapid gas inflow fluctuations or in terms of longer-term variations in galaxies' gas reservoirs \citep[e.g.][]{mannucci2010,andrews2013,dayal2013,kashino2016}. However, recent works have found equivalent `fundamental' relations for the gaseous N/O ratio \citep{hp2022} and for galaxies' stellar metallicity \citep{faisst2016, looser2024,zhuang2024}, with both relations extending up to higher stellar masses than the traditional FMR \citep{boardman2025}. Nitrogen is released over longer ($\mathrm{>100~Myr}$) timescales than oxygen at a metallicity-dependent rate \citep[e.g.][]{edmunds1978,matteucci1986,molla2006,vincenzo2016,johnson2023}, with N/O thus expected to be highly sensitive to galaxy SFHs \citep[e.g.][]{johnson2023,boardman2024,boardman2024a,pilyugin2024} while possibly less sensitive to sudden metal-poor gas inflows \citep{kashino2016,luo2021}. Stellar metallicity, meanwhile, is effectively a weighted history of gas metallicity and so is also sensitive to SFHs \citep[e.g.][]{frasermckelvie2022} and to chemical evolution histories. It is therefore likely that the FMR is driven by processes that operate over longer ($\gtrsim$ Gyr) timescales than is sometimes assumed, as for instance argued by \citet{looser2024}. Residual metallicity trends between gas metallicity and dark halo mass \citep{baker2023a, yang2024} likewise suggest that metallicities vary over long timescales.   

Recent years have seen renewed interest in the size--metallicity connection. \citet{deugenio2018} demonstrated that $\mathrm{M_*/R_e}$ (hereafter $\Phi_e$), where $\mathrm{R_e}$ is half-light radius, correlates especially tightly with gas metallicity for galaxies within the Sloan Digital Sky Survey (SDSS). \citet{sm2024} reached similar conclusions for Mapping Nearby Galaxies at Apache Point Observatory (MaNGA) survey galaxies, finding $\Phi_e$ to be the most predictive of metallicity out of over 100 tested parameters (including $\mathrm{M_*}$) via a random forest regression analysis. \citet{ma2024a} show $R_e$ to be significantly correlated with MZR residals, with SFR showing a far milder relation. Metallicity can be predicted even more tightly by considering $\mathrm{R_e}$ and SFR together with $M_*$ \citep{ma2024a,boardman2025}.

$\Phi_e$ has previously been considered as a proxy for gravitational potential \citep[][]{deugenio2018,ma2024a,sm2024}. In this view, the tight $\Phi_e$-metallicity relation can be explained as being directly due to the effects of potential, with higher potentials conferring greater resistance to outflows and so leading to higher equilibrium metallicities. However, multiple problems have emerged with this view. Metallicity anti-correlates with central dynamical mass at fixed $\mathrm{M_*}$ in observations \citep{baker2023a}, which should not occur if potential drives metallicity. In addition, metallicities in the Evolution and Assembly of GaLaxies and their Environments (EAGLE) cosmological hydrodynamical simulations show no connection with escape velocity \citep{sm2018a}, further suggesting that potential is not what truly drives metallicity. 

An alternative view of $\Phi_e$ is that it serves as a proxy for galaxies' broad SFH shapes \citep{boardman2025}, as evidenced by its especially tight correlation with N/O \citep{boardman2024a}. $\Phi_e$ is also closely related to the stellar metallicities of both star-forming and quiescent galaxies \citep{barone2018,barone2020,vaughan2022}, supporting $\Phi_e$ as an indicator of SFHs and of chemical evolution histories. All of this suggests the $\Phi_e$--metallicity connection to be due to $\Phi_e$ linking closely to galaxies' broad evolution histories, which themselves then link closely to galaxies' chemical abundances.

Given all of the above, two parallel views have developed concerning gaseous chemical abundances in galaxies. In one view, MZR scatter is driven by the current status of galaxies' gas reservoirs, with the well-known FMR emerging as a projection. In the other view, MZR scatter is driven by galaxies' gravitational potentials or else by their long-term SFHs, with $\Phi_e$ acting as a proxy in both cases. These views should in practice be connected: gas masses display positive correlations with stellar galaxy sizes at fixed stellar mass \citep{pan2021}, with gas masses also correlating with other optical galaxy properties including color, surface density and light concentration \citep{bothun1984,kannappan2004,zhang2009,wu2020,li2022}. Thus, we address the two views together for the first time, using a sample of star-forming galaxies in the nearby ($z \lesssim 0.15$) Universe. We will show that the MZR scatter relates most fundamentally to the mass of gas within galaxies' optical radii, with galaxy optical sizes acting as a close proxy for this effect. With a series of chemical evolution models, we will then suggest that our results can be explained in large part from broad variations in galaxies' inflow histories, with earlier inflow histories corresponding to lower later-time gas masses along with higher metallicities at fixed $\mathrm{M_*}$. 

We will also address the meaning of $\Phi_e$ in this article, by assessing galaxy dynamical masses across the stellar mass-size plane. We will show that more extended galaxies possess larger dynamical masses at fixed $\mathrm{M_*}$, suggesting that $\Phi_e$ is \textit{not} indicative of gravitational potential. Rather, $\Phi_e$ simply indicates the concentration of galaxies' stars, making it a powerful proxy for SFHs and of gas consumption histories in turn. We therefore advocate the phrase `stellar concentration' to describe this parameter.

The layout of this paper is as follows. In Section \ref{sampledata}, we present the sample and data employed in our analysis. In Section \ref{analysis} we describe the statistical methods employed in this work. We present our observational results in Section \ref{results}, before presenting comparisons with chemical evolution models in Section \ref{chemevo}. We discuss our findings in Section \ref{disc}, before summarizing and concluding in Section \ref{summary}. We assume a standard $\Lambda$CDM cosmology throughout ($\mathrm{h = 0.71, \Omega_m = 0.27, \Omega_\lambda = 0.73}$), along with a \citet{kroupa2001} IMF.

\section{Sample \& data}\label{sampledata}

We analyzed star-forming galaxies from the SDSS-IV \citep{blanton2017} MaNGA survey \citep{bundy2015}. As of SDSS data release 17 \citep[DR17;][]{sdssdr17}, all MaNGA data and analysis products are publically available. The main MaNGA survey observed $\sim$10000 nearby galaxies ($z \lesssim 0.15$) with a roughly flat log(mass) distribution and a wide range of sizes, morphologies, inclinations and environments \citep{yan2016b}. Roughly 2/3 of the galaxies were observed out to $\sim$1.5 $\mathrm{R_e}$ along their major axis, with these galaxies comprising MaNGA's primary and color-enhanced samples \citep{wake2017}; the remaining 1/3 of galaxies, which form MaNGA's secondary sample, were instead observed out to $\sim$2.5 $\mathrm{R_e}$. MaNGA observations were taken using the BOSS spectrographs on the 2.5 m Sloan telescope at Apache point observatory \citep{gunn2006}; these observations employed hexagonal optical fibre bundles numbering between 19 and 128 individual fibres \citep{drory2015}. The observations employed a three-point dither pattern to fully sample the field of view \citep{law2015}; they were reduced using the MaNGA data reduction pipeline \cite[DRP;][]{law2016,yan2016a,law2021} and subsequently analysed with the MaNGA data analysis pipeline \cite[DAP;][]{belfiore2019,westfall2019}. DRP and DAP products can be retrieved through the \textsc{marvin} interface \citep[][]{cherinka2019}, which is available online and as a \textsc{python} package\footnote{\url{https://www.sdss4.org/dr17/manga/marvin/}}. 

We describe our main galaxy sample --- consisting of star-forming HI-detected galaxies with metallicity measurements at 1~$\mathrm{R_e}$ --- in Section \ref{mainsamp} along with describing our obtained galaxy properties. We describe indirect estimates of galaxies' molecular gas masses in Section \ref{molmass}. We then describe a subsample with reliable dynamical masses ($\mathrm{M_{Dyn}}$) in Section \ref{mdynsamp}, which we will use to investigate central dynamical masses across the mass-size plane.  

\subsection{Main sample}\label{mainsamp}

We draw our main sample from two MaNGA value-added catalogs: the \textsc{pypipe3d} summary catalog \citep{sanchez2016,sanchez2016b,sanchez2018,sanchez2022} and the DR3 release of the HI-MaNGA catalog \citep{masters2019,stark2021}. We also use various properties from the NASA-Sloan-Atlas (NSA) catalog \citep{blanton2011}, which is based on SDSS photometry; these properties are the elliptical petrosian stellar mass ($\mathrm{M_*}$), the r-band elliptical Petrosian half-light radius ($\mathrm{R_e}$) and the elliptical Petrosian axis ratio ($\mathrm{b/a}$). We specifically choose NSA values for $\mathrm{M_*}$ and $\mathrm{R_e}$ over \textsc{pypipe3d} equivalents because of \textsc{pypipe3d} values being calculated within the MaNGA field of view. The NSA catalog does not provide errors for $\mathrm{M_*}$ or $\mathrm{R_e}$; we therefore assume constant measurement errors of 0.1 dex and 0.05 dex for $\mathrm{M_*}$ and $\mathrm{R_e}$ respectively, motivated by typical uncertainties on these parameters for nearby galaxies \citep[e.g.][and references therein]{deugenio2018,frasermckelvie2019b}.

\subsubsection{\textsc{pypipe3d}}

We refer readers to \citet{sanchez2022} and references therein for a detailed description of \textsc{pypipe3d}'s procedures, with a brief summary provided here. For deriving stellar population properties, \textsc{pypipe3d} bins spaxels for increased S/N while preserving the isophotal shape as much as possible \citep{sanchez2016b}; it performs spectral fits with 273 simple stellar population (SSP) spectra derived from the MaStar stellar library \citep{yan2019}, with the spectra covering an age range of 1~Myr to 13.5~Gyr and a metallicity range of $0.006 \leq Z/Z_\odot \leq 2.353$. These fits include dust attenuation corrections using a \citet{cardelli1989} extinction law with $\mathrm{R_V = 3.1}$. Gas metallicities are determined spaxel-by-spaxel with a wide range of strong emission calibrations, with emission fluxes dust-corrected by assuming an intrinsic H~$\alpha$ to H~$\beta$ ratio of 2.86; this procedure is restricted to star-forming spaxels which are selected using BPT diagnostics for $\mathrm{[OIII]/H\beta}$ and $\mathrm{[NII]/H\alpha}$ \citep{bpt,kewley2001} and by requiring a minimum equivalent equivalent width of 3~\AA\ for H~$\alpha$ emission lines. \textsc{pypipe3d} determines 1~$\mathrm{R_e}$ properties by averaging spaxel values within elliptical annuli and then by performing straight line fits between 0.5--2~$\mathrm{R_e}$. \textsc{pypipe3d} determines errors via Monte Carlo resimulations wherein noise is applied to spectra with the full analyses then re-ran. If a galaxy contains an insufficient amount of star-forming spaxels, then gas abundance information is not recorded for that galaxy. 

We obtain from \textsc{pypipe3d} the following chemical abundance measures at 1~$\mathrm{R_e}$: gas metallicities (O/H), nitrogen-to-oxygen ratios (N/O) light-weighted stellar metallicities ($\mathrm{[Z/H]_*}$) and light-weighted ages ($\mathrm{t_{LW}}$), with the latter motivated by past reports of residual age--metallicity relationships \citep[e.g.][]{lian2015,dp2022} \rev{along with known anti-correlations between size and stellar age at fixed $\mathrm{M_*}$ \citep{scott2017,barone2020,robotham2022}}. We also obtain SFRs from \textsc{pypipe3d}, using values obtained by summing the dust-corrected H~$\alpha$ emission flux over the full MaNGA field of view (\textit{log\_SFR\_Ha}), along with obtaining light-weighted stellar ages at 1$\mathrm{R_e}$. Our adopted SFRs are calculated by summing emission across all spaxels in a given galaxy, but we have verified that restricting to star-forming spaxels (\textit{log\_SFR\_SF} in \textsc{pypipe3d}) produces no significant difference. Galaxy properties at 1~$\mathrm{R_e}$ are tightly indicative of global stellar and gaseous properties \citep{gonzalezdelgado2014,gonzalezdelgado2015,sanchez2016b} while also being resistant to aperture effects, which is a significant concern for single-fibre studies of galaxy metallicities  \citep[e.g.][]{deugenio2018}. Throughout this article, we will focus on gas metallicities measured with the \citet{curti2020} R23\footnote{$\mathrm{R23 = ([OII]_{3727,3729}+[OIII]_{5007})/H\beta}$} calibrator (the \textit{OH\_Cur20\_R23\_Re\_fit} column in \textsc{pypipe3d}). We favor the R23 calibration due to it making no implicit assumptions as to galaxies' abundance ratios, which is a limitation of many commonly-applied calibrators \citep[e.g.][]{maiolino2019}, and due to R23 possessing relatively little intrinsic metallicity scatter \citep{curti2020}; however, we show in Appendix~\ref{altcallib} that we obtain very similar results with alternative calibrators. For N/O, we use values derived from equation 16 of \citet{pilyugin2016}, which estimates N/O from a combination of $\mathrm{[OII]_{3727,3729}}$, $\mathrm{[NII]_{6548,6584}}$ and H~$\beta$ fluxes (the \textit{NO\_Pil16\_N2\_R2\_Re\_fit} column in \textsc{pypipe3d}).

We first selected from \textsc{pypipe3d} a sample of massive ($\mathrm{M_* \geq 10^{8.8}M_\odot}$) non-edge-on ($\mathrm{b/a} > 0.4$) galaxies with 1~$\mathrm{R_e}$ gas metallicity measurements and with metallicity errors below 0.1~dex. We further restricted to galaxies with SFRs no more than 1~dex below the fitted star-forming sequence from \citet{saintonge2022}. We removed one galaxy with an anomalously high SFR which, on inspection of its central spectrum, appeared to be AGN-contaminated. We removed one galaxy with a metallicity on the lower R23 branch, since we expect massive galaxies to lie on the upper branch. We removed four other galaxies with anomalous low metallicity values ($\mathrm{12 + \log(O/H) < 8.15}$), wherein metallicity is poorly constrained for our chosen calibrator \citep{curti2020}. We confirmed that all selected galaxies also contain recorded values for $\mathrm{\log(N/O)_e}$ and $\mathrm{[Z/H]_e}$. This resulted in 3673 galaxies being selected.

\subsubsection{HI-MaNGA}

Following our \textsc{pypipe3d} selections, we cross-matched the resulting sample with HI-MaNGA. We imposed further selection criteria at this stage: we required a HI detection with $\mathrm{S/N > 3}$, with no likely source confusion (CONFLAG $= 0$) and no corruption from negative signal (NEGDET~$= 0$) or from strong baseline variations (BLSTRUCT~$= 0$). This resulted in a final sample of 1542 objects. We obtained neutral hydrogen masses ($\mathrm{M_{HI}}$) from HI-MaNGA; HI-MaNGA does not directly provide errors for $\mathrm{M_{HI}}$, so we estimated these from the errors provided on the HI flux.

\citet{wang2016} showed nearby galaxies to follow a near-uniform outer HI profile when scaling by HI disk size ($\mathrm{r_{HI}}$), which specifies the radius of a HI disk defined at a surface density of 1~$\mathrm{M_\odot/pc^2}$. \citet{wang2020} subsequently used this to estimate `inner' HI masses ($\mathrm{M_{HI,in}}$) within galaxies' r-band 90\% light radii ($\mathrm{R_{90}}$), with \citet{chen2022a} then showing $\mathrm{M_{HI,in}}$ to relate very closely to MaNGA galaxy metallicities at fixed $\mathrm{M_*}$. Motivated by this, we also estimate $\mathrm{M_{HI,in}}$ for our sample galaxies using the method of \citet{wang2020}, using r-band $\mathrm{R_{90}}$ from the NSA catalog. We estimate $\mathrm{r_{HI}}$ using the HI mass-size relation presented in \citet{wang2020}, which we use to scale the \citet{wang2016} average HI profile\footnote{We extracted this from figure 1 of \citet{wang2020}, using the web plot digitiser at \url{https://web.eecs.utk.edu/~dcostine/personal/PowerDeviceLib/DigiTest/index.html}} for each sample galaxy. We integrate the resulting profiles between $\mathrm{R_{90}}$ and 1.5~$\mathrm{r_{HI}}$ to obtain `outer' HI masses, which we then subtract from $\mathrm{M_{HI}}$ to obtain $\mathrm{M_{HI,in}}$. As in \citet{wang2020}, we assume $\mathrm{M_{HI,in} = M_{HI}}$ in cases where $\mathrm{R_{90} > 1.5r_{HI}}$. We determine errors on each $\mathrm{M_{HI,in}}$ by running 50 Monte Carlo resimulations, with Gaussian random noise added to $\mathrm{M_{HI}}$ and to $\mathrm{R_{90}}$. We assume errors of 0.05 dex for $\mathrm{R_{90}}$, motivated by typical errors on galaxy sizes \citep{deugenio2018} along with the NSA catalog's lack of provided $\mathrm{R_{90}}$ errors. 

\subsection{Molecular and total gas masses}\label{molmass}

We estimate molecular gas masses ($\mathrm{M_{mol}}$) for the galaxies in our sample by employing the tight relationship between specific star-formation rate ($\mathrm{sSFR = SFR/M_*}$) and $\mathrm{M_{mol}/M_*}$ observed in the extended CO Legacy Database for the GALEX\footnote{Galaxy Evolution Explorer} Arecibo SDSS Survey \citep[xCOLD GASS;][]{saintonge2017}. Specifically, we use the binning-derived sSFR-$\mathrm{M_{mol}/M_*}$ relation for the full xCOLD GASS sample, presented in table 5 of \citet{saintonge2017}. We perform a straight-line fit to this relation, and we estimate $\mathrm{M_{mol}}$ for our galaxies from their SFRs and stellar masses. We determine errors on $\mathrm{M_{mol}}$ via 100 Monte-Carlo resimulations, with Gaussian noise added to SFR and $M_*$ as well as to the bins of the \citet{saintonge2017} relation.

We estimate our galaxies' total gas masses as $\mathrm{M_{gas} = M_{HI} + M_{mol}}$. We also define total \emph{inner} gas masses as $\mathrm{M_{gas,in} = M_{HI,in} + M_{mol}}$. We remind readers that our SFRs were determined within the MaNGA FOV, meaning that $\mathrm{M_{HI,in}}$ and $\mathrm{M_{mol}}$ are estimated over similar spatial scales.

\subsection{$\mathrm{M_{Dyn}}$ subsample}\label{mdynsamp}

From the above-described galaxy sample, we constructed a subsample with reliable dynamical mass estimates by cross-matching with the MaNGA DynPop catalog of \citet{zhu2023}, who constrain dynamical properties using the Jeans Anisotropic Mass (JAM) modelling method \citep{cappellari2008, cappellari2020}. We specifically use HDU4 of the \citet{zhu2023} catalog, which presents results from cylindrical JAM models with NFW \citep{nfw} dark halos. Dynamical masses in this catalog are calculated within 1~$\mathrm{R_e}$. As per the recommendations of \citet{zhu2023}, we restricted to galaxies with visual fit quality flags of 1 or greater; this ensures that the subsample only contains galaxies with reliable dynamical masses. The resulting subsample, which we refer to hereafter as the $\mathrm{M_{Dyn}}$ subsample, contains 691 galaxies.

\subsection{Sample summary}

To summarize, we selected a sample of star-forming MaNGA galaxies with reliable chemical abundance measurements and reliable HI masses, which we refer to as our main sample. We estimated molecular gas masses for these galaxies using their specific star-formation rates, which allowed us in turn to estimate total gas masses. We also selected a subsample which additionally had reliable dynamical masses within 1~$\mathrm{R_e}$, which we refer to as our $\mathrm{M_{Dyn}}$ subsample. 

\begin{figure}
\begin{center}
	\includegraphics[trim = 0.2cm 0.4cm 0cm 0cm,scale=0.63]{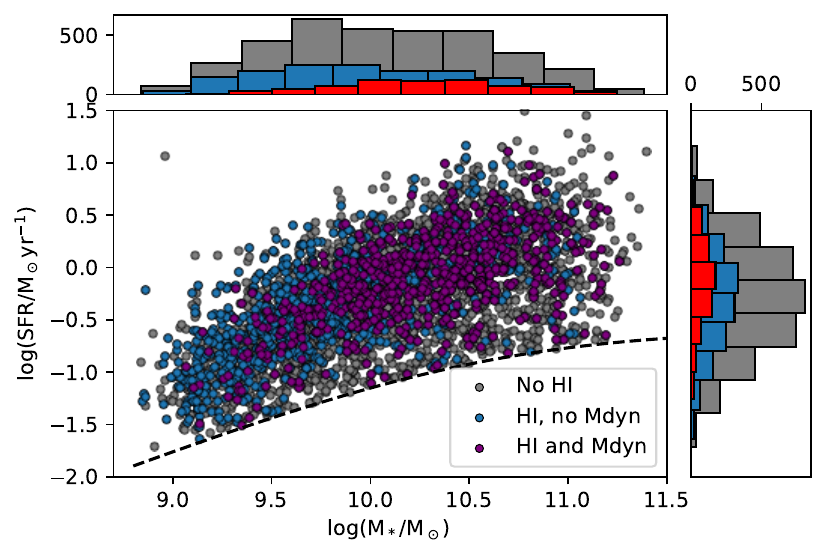}  
	\caption{Star-forming sequence (SFS) for our star-forming galaxy samples. Blue points indicate galaxies in our main sample, while purple points indicate galaxies that are also within the $\mathrm{M_{Dyn}}$ subsample. In the histograms, blue bars indicate the main sample and the red bars the $\mathrm{M_{Dyn}}$ subsample. We also show galaxies without reliable HI masses with gray points and bars. The dashed line indicates where SFR reaches 1 dex below the star-forming sequence of \citet{saintonge2022}, which we required our samples to be above.}
	\label{sampleplot}
	\end{center}
\end{figure}

We plot our resulting samples in terms of $\mathrm{M_*}$ and SFR in Figure \ref{sampleplot}, while also showing in gray the galaxies without reliable HI. By construction, our samples present a tight star-forming sequence (SFS). Our HI cut preferentially removes galaxies with less star-formation, though the SFS remains well-sampled, while the $\mathrm{M_{Dyn}}$ subsample preferentially contains high-mass galaxies.  

\section{Data analysis methods}\label{analysis}

In the following subsections we describe the statistical methods that we will employ in our analysis.

\subsection{Two-dimensional smoothing}

Over much of this paper, we will plot a parameter as a function of two other parameters. In most cases, to more clearly see two-dimensional trends, we employ two-dimensional locally weighted regression (LOESS) smoothing \citep{cleveland1988} as implemented in \textsc{python} \citep{cappellari2013a}\footnote{\url{https://users.physics.ox.ac.uk/~cappellari/software/\#sec:loess}}. For a given data point, we compute its LOESS-smoothed value using the closest 10\% of datapoints in a given parameter space with the \textit{rescale} keyword applied. For weighting purposes, the LOESS algorithm computes errors from the scatter in neighbouring points.

The LOESS method is useful only for data visualisation. For quantitative analysis, we use partial correlation coefficients and random forest (RF) regression as detailed below. For interested readers, we also provide smoothing-free versions of plots in Appendix~\ref{nosmooth}.

\subsection{Partial correlation coefficients}

Partial correlation coefficients ($\rho_{ij,k}$) describe the strength of correlation between two parameters $i$ and $j$ once a third parameter $k$ is controlled for. Partial correlation coefficients have been employed in various recent galaxy studies \citep[e.g.][]{bait2017,bluck2020,baker2023,boardman2024a}, and are calculated from Spearman rank correlation coefficients using

\begin{equation}
\rho_{ij,k} = \frac{\rho_{ij}-\rho_{ik}\rho_{jk}}{\sqrt{1-\rho^2_{ik}}\sqrt{1-\rho^2_{jk}}}
\end{equation}

\noindent
where $\rho_{ij}$ denotes the Spearman correlation between any two parameters. We use combinations of $\rho_{ij,k}$ to compute the angle of maximum increase along a parameter space, using

\begin{equation}
\tan(\theta_{k,ij}) = \left(\frac{\rho_{jk,i}}{\rho_{ik,j}}\right)
\end{equation}
\noindent
where $\theta_{k,ij}$ describes the angle of maximum increase for a parameter $k$ in parameter space $i,j$, \rev{with $\theta_{k,ij}$ calculated counter-clockwise in the positive $i$ direction.}

The main limitation of partial correlation coefficients is that they can only consider two parameters besides metallicity at a time; thus, when using this method, we implicitly assume that we are considering the most appropriate physical parameters for determining metallicity. 

\subsection{Random forest regression}

To rectify the above-mentioned limitation, we also employ RF regression \citep{breiman2001}. RF regression is a method of predicting an output parameter from a set of input parameters which are termed \textit{features}; in doing this, a random forest regressor determines the relative importance of different features for making the prediction. RFs have seen significant use in recent galaxy metallicity studies on resolved and global scales \citep[e.g.][]{sm2019,baker2023a,baker2023b,sm2024,sm2024a,koller2025,lyu2025}, motivated by the large number of observed metallicity scaling relations, with RFs also used widely across extragalactic astronomy \citep[e.g.][]{carliles2010,clarke2020,mucesh2021,bluck2022}. 

Throughout this article, we employ the \textsc{python} implementation of the RF method provided by \textsc{scikit-learn} \citep{scikit-learn}. We provide a brief overview of the RF method along with detailing our implementation. For a more in-depth description of the RF method, we refer interested readers to Scikit-learn's documentation\footnote{\url{https://scikit-learn.org/stable/modules/ensemble.html\#forest}}.

The RF method is an extension of the decision tree method, which itself is purely deterministic. In decision tree regression, a target variable is predicted through a series of binary splits on a single feature, with splits --- in terms of the chosen feature and the value to split on --- determined so as to minimise a chosen impurity measure. Splits are performed on `nodes', with `leaf nodes' containing the final groupings and predictions. The RF method extends this by splitting a supplied dataset into random subsamples on which a decision trees are constructed. For each data point, the RF prediction is then the average from all decision trees.

For our RF implementation, we use the mean-square error (MSE) as our impurity measure and we consider all features for each split as opposed to randomly selecting a subset. We split our full galaxy dataset into training and test sets, utilizing a 75--25 split in the sets' sizes. The RF regressor is fitted to the training set, with the test set then used to verify the regressor's accuracy.

We run the RF with the following hyperparameters: $\mathrm{n_{estimators} = 1000}$, $\mathrm{max_{depth} = 7}$, $\mathrm{min_{split} = 24}$, $\mathrm{min_{leaf} = 12}$. Here, $\mathrm{n_{estimators}}$ describes the number of individual trees, $\mathrm{max_{depth}}$ the maximum number of tree layers, $\mathrm{min_{split}}$ the minimum size of a node that can be split, and $\mathrm{min_{leaf}}$ the minimum allowed size of a leaf node. We tuned these hyperparameters by hand, aiming to maximize model performance while also achieving similar performances on the training and test sets; this latter point is to ensure that we do not overfit the training set. We have also verified that our results are not significantly altered if different hyperparameter combinations are chosen.

As in other recent metallicity works \citep{baker2023a,baker2023b,sm2024,sm2024a,koller2025}, we compute impurity-based importances to determine the features most closely connected to metallicity. These importances are computed on the training set as the mean decrease in impurity (MDI) achieved by a given feature across all trees in the random forest, with the MDI then normalised to a total of 1. An alternative method is to measure permutation importances, which are computed on a separate test set by randomly reshuffling a feature and then measuring the resulting impurity increase; this is less suitable for our purposes, due to our RF implementation including heavily correlated features, so we prefer impurity importances for our analysis. We \rev{have verified} however that using permutation importances produces no significant difference in our findings.

Following \citet{sm2024}, we perform 50 RF realisations with different random training/test splits in each case. We then report feature importances as the mean MDI across all realisations, with errors reported as standard deviations.

\section{Observational results}\label{results}

We begin here by investigating galaxy dynamical masses across the stellar mass--size plane (Section \ref{res2}), to better understand the meaning of $\mathrm{\Phi_e}$. We will then investigate gas masses across the mass-size plane (Section \ref{res1}), noting in particular the tight correlation between $\mathrm{M_{HI,in}}$ and $\mathrm{R_e}$. Finally, we investigate metallicity scaling relations in Section \ref{res3}, employing detailed statistical analyses to address the driving factors behind gas metallicities. 

\subsection{Dynamical masses across the $M_*$--$\mathrm{R_e}$ plane and the meaning of $\Phi_e$}\label{res2}

Various previous works have shown a tight correlation between $\Phi_e$ and metallicity, with this correlation most commonly interpreted as being driven by gravitational potential \citep{deugenio2018,sm2024,ma2024a,koller2025}. \citet{baker2023a} have however challenged this interpretation, pointing out an inverse trend between $\mathrm{M_{Dyn}}$ and gas metallicity at fixed $\mathrm{M_*}$. Thus, to understand this situation further, we consider both metallicity and $\mathrm{M_{Dyn}}$ across the galaxy mass-size plane.

\begin{figure}
\begin{center}
    \includegraphics[trim = 0.06cm 0.3cm 0cm 0cm,scale=0.75]{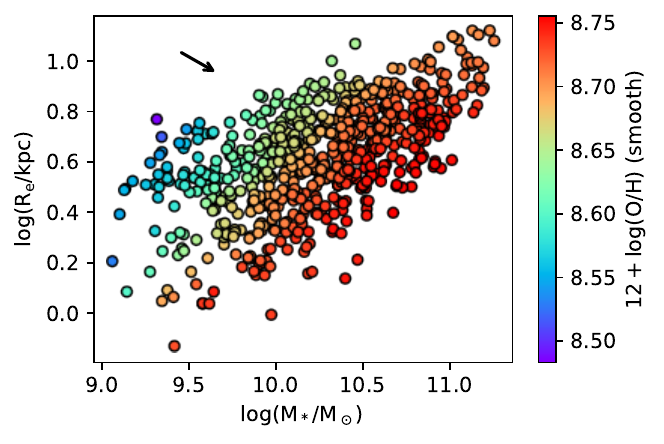} 
	\includegraphics[trim = 0cm 0.5cm 0cm 0cm,scale=0.75]{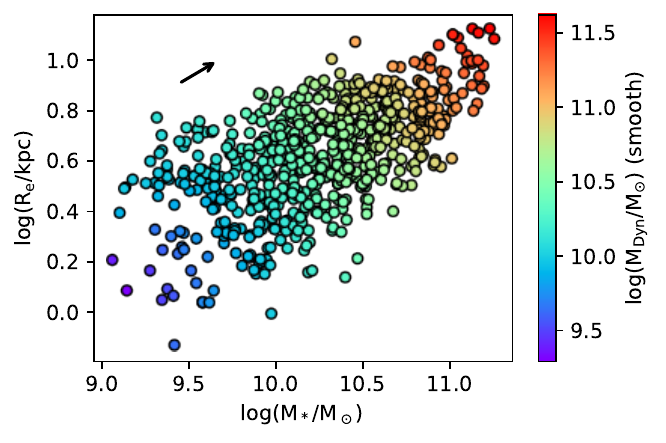} 
    \caption{Top: gas metallicity as a combined function of stellar mass ($\mathrm{M_*}$) and half-light radius ($\mathrm{R_e}$), for the $\mathrm{M_{Dyn}}$ subsample with smoothing applied. The arrow shows the direction of maximum metallicity increase and is computed from partial correlation coefficients. Bottom: as above, but for $\mathrm{M_{Dyn}}$. We find $\mathrm{M_{Dyn}}$ to rise with both increasing $\mathrm{M_*}$ and increasing $\mathrm{R_e}$. This suggests that $\mathrm{\Phi_e}$ ($\mathrm{=M_*/R_e}$) is \emph{not} a good proxy for the gravitational potential depth, and \rev{that} potential does not drive the tight $\mathrm{\Phi_e}$--metallicity relation.}
	\label{mre_dyn}
	\end{center}
\end{figure}

In Figure \ref{mre_dyn}, we present gas metallicity and $\mathrm{M_{Dyn}}$ across the stellar mass-size plane for our $\mathrm{M_{Dyn}}$ subsample, with LOESS smoothing applied. We also show the directions of maximum increase in both cases, which we compute by applying Equations 1 and 2 to the data points without smoothing. We find that $\mathrm{M_{Dyn}}$ positively correlates with both $\mathrm{M_*}$ and $\mathrm{R_e}$, such that more compact galaxies possess \emph{lower} $\mathrm{M_{Dyn}}$ for their $\mathrm{M_*}$. Combined with more compact galaxies possessing higher metallicities for their $\mathrm{M_*}$, this results in an inverse $\mathrm{M_{Dyn}}$--metallicity correlation at fixed stellar mass, in agreement with \citet{baker2023a}.

From this result, we argue that $\Phi_e$ --- defined as $\mathrm{M_*/R_e}$ --- is \textit{not} a good proxy for the depth of gravitational potential, contrary to what has previously been assumed. Instead, $\mathrm{M_*/R_e}$ simply represents the concentration of stellar mass, and we advocate the term `stellar concentration' for this parameter.

\subsection{Gas trends in mass--size space}\label{res1}

\begin{figure}
\begin{center}
	\includegraphics[trim = 0.5cm 0.1cm 0cm 0.2cm,scale=0.75]{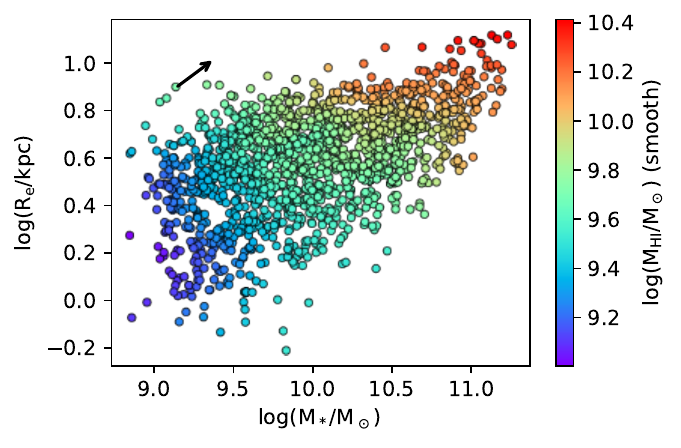} 
    \includegraphics[trim = 0.3cm 0.1cm 0cm 0.2cm,scale=0.75]{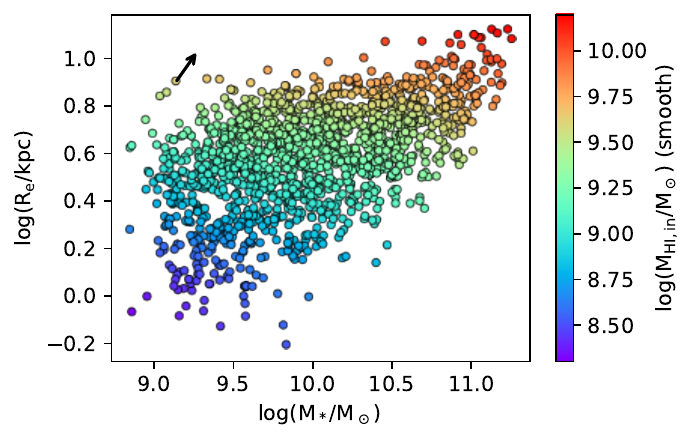} 
    \includegraphics[trim = 0.6cm 0.7cm 0cm 0.2cm,scale=0.75]{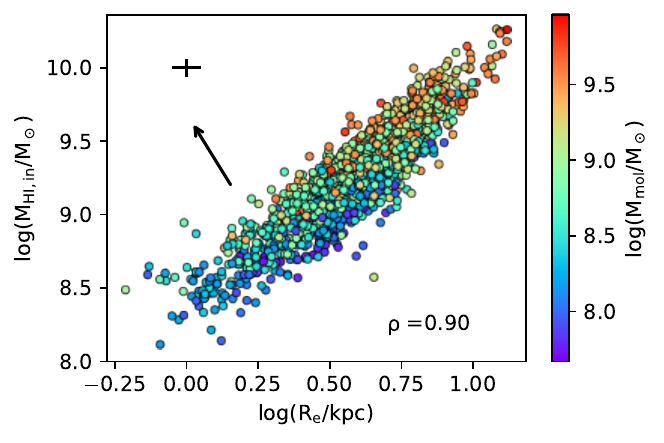} 
	\caption{Top: global HI gas mass ($\mathrm{M_{HI}}$) plotted as a combined function of stellar mass ($\mathrm{M_*}$) and half-light radius ($\mathrm{R_e}$) with LOESS smoothing applied. The arrow shows the direction of maximum $\mathrm{M_{HI}}$ increase and is computed using partial correlation coefficients. Middle: as above, but for the HI mass within galaxies' 90\% light radii ($\mathrm{M_{HI,in}}$). We find $\mathrm{M_{HI}}$ to rise with both increasing $\mathrm{M_*}$ and increasing $\mathrm{R_e}$, consistent with \citet{pan2021}, such that more extended galaxies possess greater gas mass at fixed stellar mass. $\mathrm{M_{HI,in}}$ instead varies almost entirely with $\mathrm{R_e}$, suggesting a close relationship between galaxies' sizes and their available gas reservoirs. Bottom: $\mathrm{M_{HI,in}}$ vs $\mathrm{R_e}$ colored by $\mathrm{\log(M_{mol}/M_\odot)}$, with median errors shown in the top left and the spearman correlation $\mathrm{\rho}$ between $\mathrm{M_{HI,in}}$ and $\mathrm{R_e}$ shown in the bottom right. A tight correlation is evident, with a residual anticorrelation emerging between $\mathrm{R_e}$ and $\mathrm{M_{mol}}$.}
	\label{mhismooth}
	\end{center}
\end{figure}

We now consider how gas masses relate to $\mathrm{M_*}$ and $\mathrm{R_e}$. We plot in the top panel of Figure \ref{mhismooth} the global HI mass as a two-dimensional function of $\mathrm{M_*}$ and $\mathrm{R_e}$, with smoothing applied. We also indicate the direction of maximal increase in $\mathrm{M_{HI}}$. We find $\mathrm{M_{HI}}$ to rise with both $\mathrm{M_*}$ and $\mathrm{R_e}$, such that more extended galaxies possess higher $\mathrm{M_{HI}/M*}$ ratios, which is consistent with \citet{pan2021}.

In the middle panel of Figure \ref{mhismooth}, we perform an equivalent analysis with $\mathrm{M_{HI,in}}$. We find $\mathrm{M_{HI,in}}$ to rise mostly with increasing $\mathrm{R_e}$, with only a small mass dependence evident at any given size. We present the relationship between $\mathrm{M_{HI,in}}$ and $R_e$ in the bottom panel of Figure \ref{mhismooth} along with the corresponding Spearman correlation ($\mathrm{\rho = 0.90}$)\footnote{All quoted Spearman correlation coefficients in this paper have P-values of $\mathrm{P \ll 0.01}$.}, where a tight correlation is evident. This behavior is partly by construction: \rev{$\mathrm{M_{HI,in}}$ is estimated within $\mathrm{R_{90}}$ and correlates tightly with it ($\mathrm{\rho = 0.96}$), with $\mathrm{R_e}$ correlating tightly with $\mathrm{R_{90}}$ in turn ($\mathrm{\rho = 0.95}$)}. Nonetheless, this result indicates that compact galaxies contain much less star-forming gas at fixed $\mathrm{M_*}$ \rev{within their optical extents}.

We color the data points in Figure \ref{mhismooth}'s final panel by $\mathrm{M_{mol}}$, with the direction of maximal increase also shown. It is apparent from this that $\mathrm{M_{mol}}$ anti-correlates with $\mathrm{R_e}$ when controlling for $\mathrm{M_{HI,in}}$ (indicated by the arrow pointing left). This may be a consequence of having estimated $\mathrm{M_{mol}}$ from sSFR, since compact star-forming galaxies display somewhat higher average sSFRs than their extended counterparts \citep{wuyts2011,boardman2025}. As a result, $\mathrm{M_{gas,in}}$ displays a slightly weaker correlation with $\mathrm{R_e}$ than $\mathrm{M_{HI,in}}$, though the correlation strength remains considerable ($\mathrm{\rho = 0.82}$). 

To summarise: we find that at fixed $M_*$, galaxies with more compact stellar distributions have lower HI gas masses than do more extended galaxies. This trend is particularly strong when we look at gas mass within the stellar/optical extent, with inner HI gas mass varying more strongly with galaxy size than galaxy mass. Thus, inner HI gas masses are significantly linked with the compactness of a galaxy's stars. 

\subsection{The gas reservoir as a fundamental metallicity driver}\label{res3}

\subsubsection{Tests with correlation coefficients}

We now consider which parameter after $\mathrm{M_*}$ is the most closely related to gas metallicity, following the spirit of various previous efforts \citep[e.g.][]{baker2023a,ma2024a,sm2024,koller2025}. Compared to these earlier works, the main difference is the inclusion of gas masses in our analysis. 

\begin{figure}
\begin{center}
	\includegraphics[trim = 0.4cm 0.5cm 0cm 0.1cm,scale=0.85]{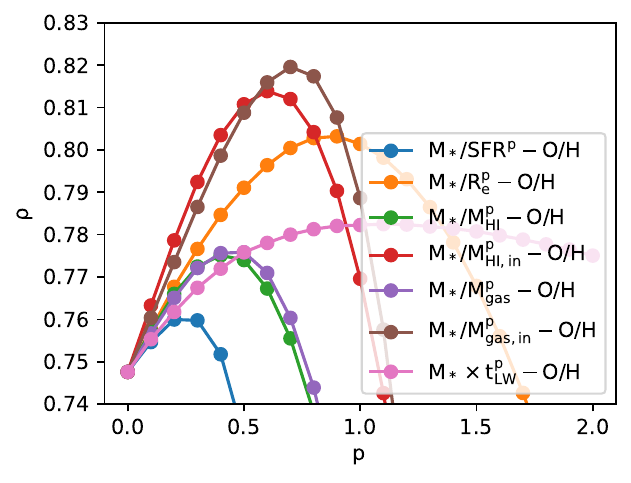} 
	\caption{Spearman correlation coefficients ($\rho$) between gas metallicity (O/H) and various parameters as indicated in the legend, with $\mathrm{p}$ representing an arbitrary power. We find $\mathrm{M_{gas,in}}$ to yield the highest $\rho$ peak and thus the strongest metallicity correlation, suggesting this parameter to be the most fundamentally connected to metallicity after the stellar mass.}
	\label{scorrs}
	\end{center}
\end{figure}

In Figure \ref{scorrs} we plot Spearman correlation coefficients between gas metallicity (O/H) and parameters of the form $\mathrm{M_*/X^{\mathrm{p}}}$, where $\mathrm{p}$ is a power between 0 and 2 that is varied to allow for slightly non-linear dependencies, and X is one of the following: SFR, $\mathrm{R_e}$, $\mathrm{M_{HI}}$, $\mathrm{M_{HI,in}}$, $\mathrm{M_{gas}}$, $\mathrm{M_{gas,in}}$. \rev{We also show correlation coefficients between O/H and $\mathrm{M_*\times t_{LW}^{\mathrm{p}}}$, motivated by previous findings of positive age-O/H correlations at fixed $\mathrm{M_*}$ \citep{lian2015,sm2020}}.  We do not show results with $\mathrm{M_{mol}}$ due to this having been determined from galaxies' sSFRs, which would cause it to functionally overlap with SFR on this figure, though we note that it produces correlations of similar strength to SFR as expected. \rev{We also don't show $\mathrm{R_{90}}$ on this figure due to its functional similarity to $\mathrm{M_{HI,in}}$ (Section \ref{res1}, second paragraph), though we note that $\mathrm{R_{90}}$ achieves very slightly weaker correlations than does $\mathrm{M_{HI,in}}$.} We obtain a number of results from this figure:

\begin{itemize}
\item Overall, $\mathrm{M_{gas,in}}$ yields the strongest metallicity correlation (as indicated by it returning the highest Spearman correlation coefficient) and so appears to be the most closely tied to metallicity out of the parameters tested. We obtain the strongest Spearman correlation ($\rho = 0.820$) when $\mathrm{p = 0.7}$. 
\item $\mathrm{M_{HI,in}}$ produces the second strongest correlation, with $\rho = 0.814$ when $
\mathrm{p = 0.6}$. 
\item When considering $\mathrm{R_e}$ specifically, the strongest metallicity correlation is achieved when $\mathrm{p = 0.9}$. This reaffirms the close connection between $\Phi_e$ (i.e. when $\mathrm{p = 1}$) and metallicity. We also find that $\Phi_e$ produces a stronger metallicity correlation than the global stellar density ($\mathrm{M_*/R_e^2}$, \rev{from comparing $\rho$ at $\mathrm{p = 1}$ with $\mathrm{\rho}$ at $\mathrm{p = 2}$}), in agreement with past work \citep{deugenio2018,sm2024}.
\item \rev{Stellar age produces a stronger maximum correlation than global gas mass measures ($\mathrm{M_{HI}}$, $\mathrm{M_{gas}}$) or SFR, though it produces a weaker maximum correaltion than do $\mathrm{R_e}$ or inner gas mass measures ($\mathrm{M_{gas,in}}$, $\mathrm{M_{HI,in}}$).}
\item SFR yields by far the weakest improvement in metallicity correlation, consistent with the FMR being only modestly less scattered than the MZR \citep[e.g.][]{salim2014,ma2024a}. As a reminder, we use SFRs derived from attenuation-corrected H~$\alpha$ fluxes \rev{within the MaNGA field of view}.
\end{itemize}

A potential concern with correlation coefficients is that larger measurement errors will artificially lower derived correlations on affected parameters. Thus, we repeated our analysis with random scatter added to $\mathrm{M_{HI}}$ and $\mathrm{R_{90}}$ (0.1 dex and 0.05 dex respectively), which resulted in noise being added to $\mathrm{M_{HI,in}}$ and hence to $\mathrm{M_{gas,in}}$. We obtained consistent results in this case, meaning that differential errors are not what drives our findings. 

\subsubsection{Alternative chemical abundance tracers}

\begin{figure}
\begin{center}
	\includegraphics[trim = 0cm 0.3cm 0cm 0.1cm,scale=0.8]{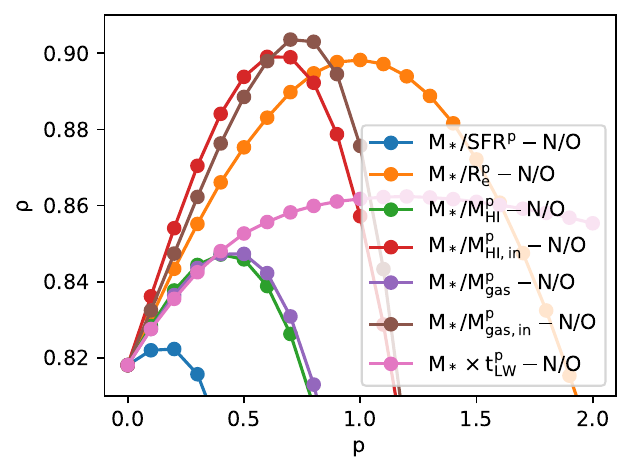}
    \includegraphics[trim = 0cm 1cm 0cm 0cm,scale=0.8]{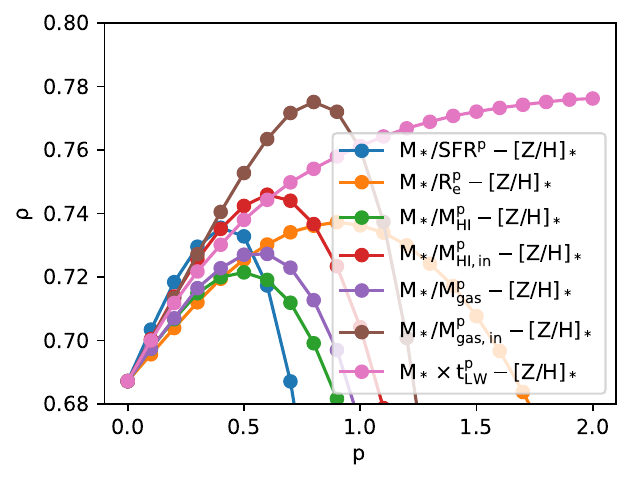}
	\caption{As in Figure \ref{scorrs}, but for N/O abundances (top) and light-weighted stellar metallicities (bottom). We find $\mathrm{M_{gas,in}}$ to produce the highest $\rho$ peak for N/O and the second-highest $\rho$ peak for stellar metallicity. The maximum N/O correlations are higher than found for O/H, in agreement with \citet{boardman2024a}.}
	\label{scorrs_no}
	\end{center}
\end{figure}

We now consider the gas-phase N/O and the stellar metallicity in the same manner as gas metallicity (O/H). We plot in the top panel of Figure \ref{scorrs_no} the Spearman correlation coefficients \rev{for N/O using the same format and the same parameters as in Figure \ref{scorrs}}. We obtain similar results as for O/H, with $\mathrm{M_{gas,in}}$ again yielding the strongest correlation and with SFR producing the weakest correlations. The maximum correlation coefficient for N/O ($\mathrm{\rho = 0.904}$) is somewhat higher than it was for O/H ($\mathrm{\rho = 0.820}$); this is consistent with \citet{boardman2024a}, where it was shown that $\Phi_e$ also correlates more closely with N/O than with O/H.

We plot in the bottom panel of Figure \ref{scorrs_no} an equivalent test for $\mathrm{[Z/H]_*}$. \rev{We find $\mathrm{t_{LW}}$ to produce the strongest correlation ($\mathrm{\rho = 0.776}$) in this case, followed closely by $\mathrm{M_{gas,in}}$ ($\mathrm{\rho = 0.775}$) and with $\mathrm{M_{HI,in}}$ producing the third strongest correlation ($\mathrm{\rho = 0.746}$). The difference in correlation strengths between $\mathrm{M_{gas,in}}$ and $\mathrm{M_{HI,in}}$, we note, is more pronounced in this case.} Given that $\mathrm{[Z/H]_*}$ is a long term average metallicity, this suggests that inner gas reservoirs are not only informative of a galaxy's \emph{current} chemical state \rev{but also} indicative of a galaxy's long-term chemical evolution history, \rev{with stellar age likewise relating to chemical evolution histories}. SFR produces relatively strong correlations here when compared to the results for O/H or N/O, which shows the FMR to be stronger for stellar than gas phase metallicity, consistent with past work \citep{looser2024,boardman2025}\footnote{The relatively strong correlation with SFR in the case of stellar metallicity may artificially boost the measured correlation coefficient with $\mathrm{M_{gas,in}}$, due to the use of SFR to compute molecular gas mass. Therefore the actual correlation coefficient with $\mathrm{M_{gas,in}}$ may in reality be closer to $\mathrm{M_{HI,in}}$.}. 

\subsubsection{Gas metallicity visualization}

\begin{figure}
\begin{center}
	\includegraphics[trim = 0.05cm 0.1cm 0cm 0.2cm,scale=0.75]{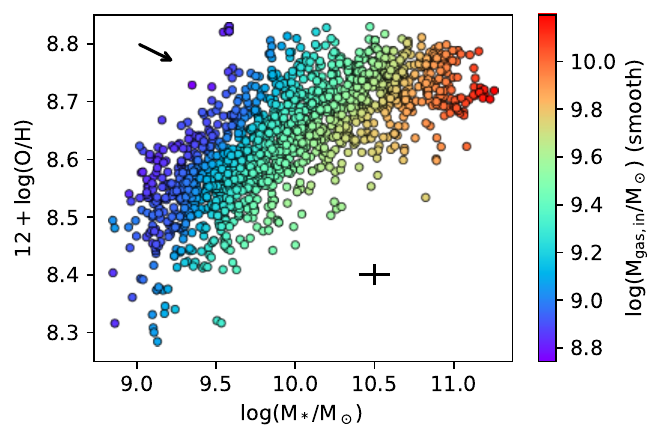} 
    \includegraphics[trim = 0.1cm 0.1cm 0cm 0.2cm,scale=0.75]{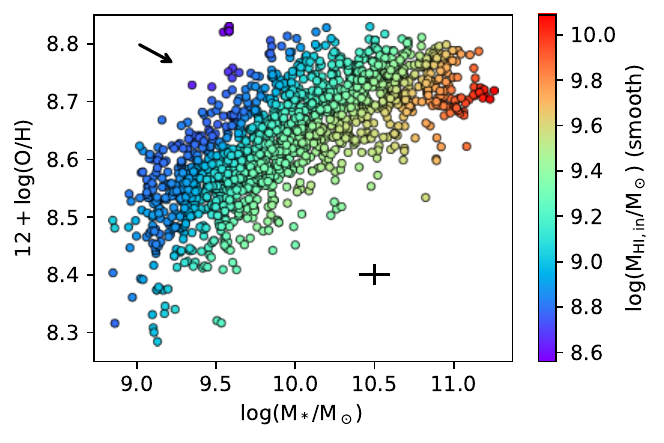} 
    \includegraphics[trim = 0.3cm 0.7cm 0cm 0.2cm,scale=0.75]{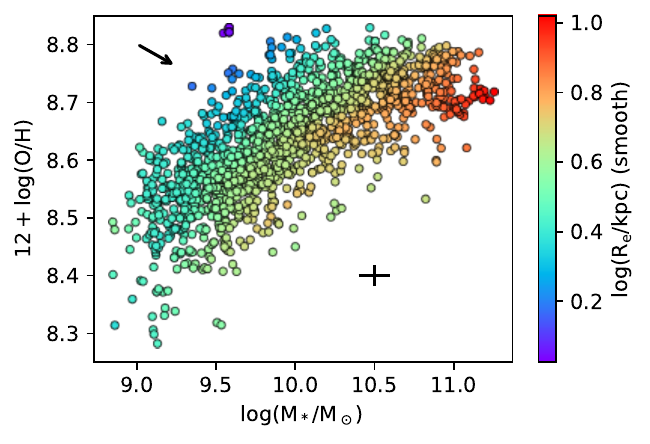} 
	\caption{Mass--metallicity relation colored by the three most constraining parameters after M$_*$ according to our correlation coefficient analysis: $\mathrm{M_{gas,in}}$ (top), $\mathrm{M_{HI,in}}$ (middle) or $\mathrm{R_e}$ (bottom), with LOESS smoothing applied in each case. The error bars show the median errors in stellar mass and metallicity, and the arrows show directions of maximum increase. We find $\mathrm{R_e}$ to move across the mass-metallicity space in a near-identical way to the gas mass measures. This suggests the mass--size--metallicity relationship to be very closely linked to that between stellar mass, gas mass and metallicity.}
	\label{msmgoh}
	\end{center}
\end{figure}

We show the mass-metallicity relation in Figure \ref{msmgoh}, with data points colored by the three most constraining parameters after $\mathrm{M_*}$ as found in Figure \ref{scorrs}: $\mathrm{M_{gas,in}}$ (top), $\mathrm{M_{HI,in}}$ (middle), and $\mathrm{R_e}$ (bottom). We find that $\mathrm{R_e}$ trends in almost the exact same direction as $\mathrm{M_{gas,in}}$ and $\mathrm{M_{HI,in}}$ across mass--metallicity space. Thus, we find that the $\mathrm{M_*}$--size--metallicity relation is intrinsically linked to the $\mathrm{M_*}$--$\mathrm{M_{HI,in}}$--metallicity relation reported in \citet{chen2022a}. To interpret these relations, we must therefore consider the relative importance of all different parameters together.

\subsubsection{Random forest analysis}

\begin{figure*}
\begin{center}
	\includegraphics[trim = 0cm 0.5cm 0cm 0cm,scale=0.85]{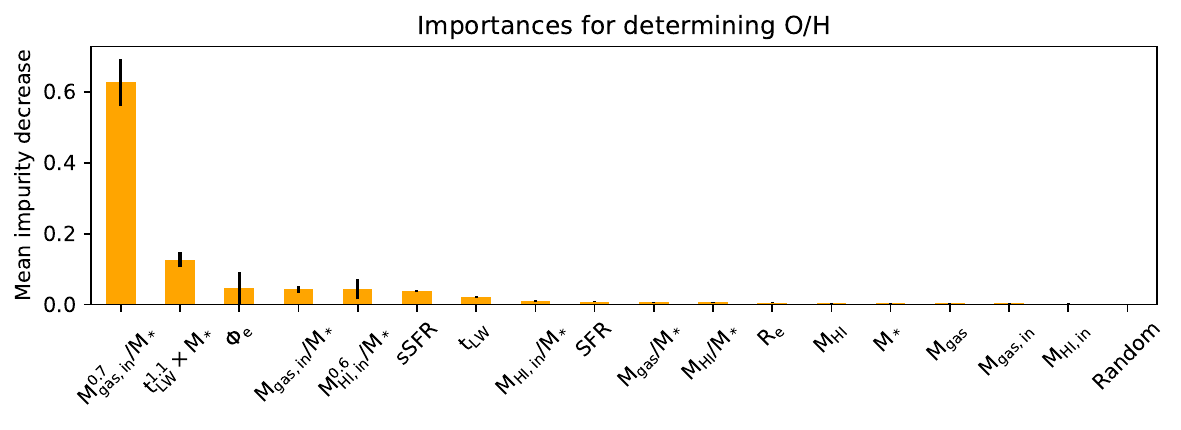} 
	\caption{Feature importances for determining gas metallicity from a random forest analysis, reported as means and standard deviations across 50 random forest realisations \rev{with importances normalised such that they sum to 1}. We find $\mathrm{M_{gas,in}^{0.7}/M_*}$ to be by far the most informative parameter for determining the metallicity. $\Phi_e$ holds little importance by comparison, despite the existence of a tight $\Phi_e$-metallicity relation.}
	\label{randomforest_main}
	\end{center}
\end{figure*}

We perform a RF analysis to determine which parameters are most closely predictive of the gas metallicity. We include the following parameters as features in our analysis: $\mathrm{M_*}$, $\mathrm{R_e}$, SFR, $\mathrm{M_{HI}}$, $\mathrm{M_{HI,in}}$, $\mathrm{M_{gas}}$, $\mathrm{M_{gas,in}}$ and light-weighted stellar age at 1~$\mathrm{R_e}$ ($\mathrm{t_{LW}}$). We also include as features some commonly-considered parameter combinations: specific star-formation rate ($\mathrm{sSFR = SFR/M_*}$), gas-to-stellar mass ratios ($\mathrm{M_{HI}/M_*}$, $\mathrm{M_{HI,in}/M_*}$, $\mathrm{M_{gas}/M_*}$, $\mathrm{M_{gas,in}/M_*}$), and $\Phi_e$. We further include $\mathrm{M_{gas,in}^{0.7}/M_*}$ $\mathrm{M_{HI,in}^{0.6}/M_*}$ \rev{and $\mathrm{t_{LW}^{1.1} \times M_*}$}, which we determined in Figure \ref{scorrs} to be the optimal projections for predicting the metallicity \rev{with their respective parameter combinations}. Finally, we include a random uniform control variable as in \citet{baker2023a} and \citet{koller2025}, to help verify that the RF is not overfitting. We obtain similar RF performance on the training and test sets, with average root-mean-square errors of 0.041 and 0.049 respectively across the 50 RF realisations; this shows that the RF is not typically overfitting the training set, which shows in turn that the RF feature importances are reliable. 

Our RF implementation includes numerous correlated features, with many of the correlations being by construction. However, this will not impact upon the RF's performance, due to our implementation considering all features in each split. As shown by the tests of \citet[][their appendix B]{bluck2022}, this kind of RF implementation can accurately determine feature importances even when features are highly correlated. 

We show in Figure \ref{randomforest_main} the feature importances from our random forest analysis. We find $\mathrm{M_{gas,in}^{0.7}/M_*}$ to be by far the most important parameter, possessing an average mean impurity decrease more than \rev{4} times higher than any other parameter, \rev{with and $\mathrm{t_{LW}^{1.1} \times M_*}$ attaining the second highest performance}. A number of parameters attain \rev{tertiary} importances that are roughly equal to each other within the errors (sSFR, $\mathrm{\Phi_e}$, $\mathrm{M_{HI,in}^{0.6}/M_*}$, $\mathrm{M_{gas,in}/M_*}$), while the remaining parameters attain little to no importance. 

\rev{The substantial difference in importances for $\mathrm{M_{gas,in}^{0.7}/M_*}$ and $\mathrm{M_{gas,in}/M_*}$ may seem surprising, given these features' close connection; however, this difference can in fact be understood as being largely \emph{because} of the features' close connection. Gas metallicity correlates more-so with $\mathrm{M_{gas,in}^{0.7}/M_*}$ (Figure \ref{scorrs}), causing the random forest to preferentially split on $\mathrm{M_{gas,in}^{0.7}/M_*}$ and leaving only a small amount of additional information to be gained from $\mathrm{M_{gas,in}/M_*}$. This results in $\mathrm{M_{gas,in}/M_*}$ attaining relatively little importance.} 

\begin{figure}
\begin{center}
	\includegraphics[trim = 0cm 1cm 0cm 0cm,scale=0.75]{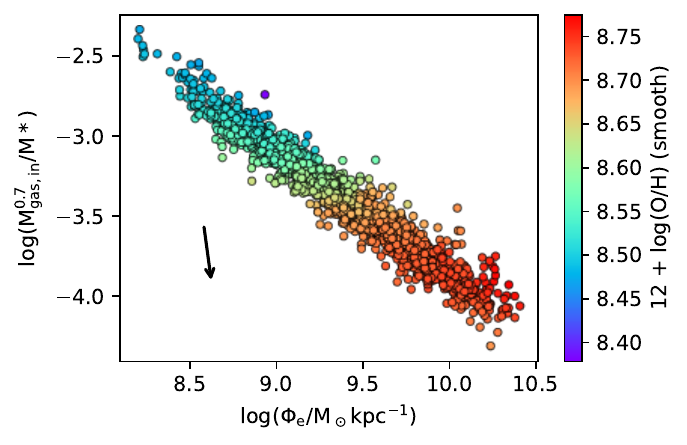} 
	\caption{$\mathrm{M_{gas,in}^{0.7}/M_*}$ vs. $\Phi_e$, colored by gas metallicity with smoothing applied. We also show the direction of maximum increase, computed from partial correlation coefficients. We find the metallicity to vary almost entirely with $\mathrm{M_{gas,in}^{0.7}/M_*}$, explaining the low importance ascribed to $\Phi_e$ by the random forest in Figure \ref{randomforest_main}.}
	\label{inmgas_phi_pcc}
	\end{center}
\end{figure}

The low importance of $\Phi_e$ is particularly interesting, given its strong reported relationship to metallicity. Figure \ref{inmgas_phi_pcc} shows the reason for this: $\Phi_e$ correlates extremely tightly with $\mathrm{M_{gas,in}^{0.7}/M_*}$, with metallicity showing almost no correlation with $\Phi_e$ at a given $\mathrm{M_{gas,in}^{0.7}/M_*}$\footnote{We note that the correlation between $\Phi_e$--$\mathrm{M_{gas,in}^{0.7}/M_*}$ will be artificially tightened by correlated errors: $\mathrm{M_*}$ is included in both parameters, and size errors will lead to correlated errors in $\mathrm{M_{HI,in}}$ which are then reflected in $\mathrm{M_{gas,in}}$. However, the main driver is the intrinsic link between HI gas mass and galaxy concentration (Section \ref{res1}).}. This behavior is reflected in the partial correlation coefficients, which are -0.29 and 0.03 when controlling for $\Phi_e$ and for $\mathrm{M_{gas,in}^{0.7}/M_*}$ respectively. Since metallicity correlates more closely with $\mathrm{M_{gas,in}^{0.7}/M_*}$, the random forest's decision trees only rarely split in $\Phi_e$, \rev{similarly to the situation discussed above between $\mathrm{M_{gas,in}^{0.7}/M_*}$ and $\mathrm{M_{gas,in}/M_*}$}. We explore this situation further in Appendix ~\ref{rfreduced}.

\emph{Overall, our RF analysis supports $\mathit{M_{gas,in}}$ as the most important parameter after $\mathit{M_*}$ in setting gas metallicities}, with other proposed secondary parameters (stellar age, SFR, size) possessing far less importance by comparison. The mass--size--metallicity relation, along with the FMR, appears to be \rev{largely} a projection of a far more fundamental relation between stellar mass, inner gas mass and gas metallicity. Furthermore, inner gas masses ($\mathrm{M_{gas,in}}$ yield tighter metallicity relations than total gas masses ($\mathrm{M_{gas}}$ or $\mathrm{M_{HI}}$), reflecting the importance of the \emph{accessible} gas supply within the galaxy disk.

\section{Chemical evolution modeling}\label{chemevo}

From our observational analysis we have established a particularly close connection between inner gas mass ($\mathrm{M_{gas,in}}$) and the chemical enrichment of galaxies, with lower $\mathrm{M_{gas,in}}$ corresponding to increased chemical enrichment at fixed stellar mass (Figure \ref{scorrs}, Figure \ref{msmgoh} top panel, Figure \ref{randomforest_main}). We now employ one-zone chemical evolution models to better understand these observed trends, using the VICE framework described in \citet{johnson2020}. 

We refer readers to \citet{johnson2020} and to the online documentation\footnote{\url{https://vice-astro.readthedocs.io/en/latest/}} for a detailed description of VICE. In brief, VICE assumes a region's gas metallicity to be driven by a combination of gas inflow, gas outflow and metal production from stars \citep[e.g.][]{lilly2013,forbes2014,wang2025}. It assumes outflows to be driven entirely by star-formation, with outflow rates ($\mathrm{\dot M_{out}}$) governed by an input mass-loading factor ($\mathrm{\epsilon = \dot M_{out}/SFR}$). Starting from standard equations, VICE numerically determines the chemical evolution of a zone (a cloud of gas with stars and uniform metallicity) based on a user-supplied SFH, inflow history, or gas mass history. The gas consumption timescale ($\mathrm{t_c}$, defined as the gas reservoir mass divided by SFR), along with $\mathrm{\epsilon}$, can be constant or time-varying. VICE considers elemental yields from core-collapse supernovae (CCSNe), type Ia supernovae (SNIa) and AGB stars. As is standard for this type of model, VICE does not consider mergers; thus, for the purpose of interpretation, a VICE model's history should be treated as being the sum of a galaxy's progenitors. 

We discuss the rationale behind our VICE models in Section \ref{vice_rationale}, describe the model generation in Section \ref{vice_generation}, and present our findings in Section \ref{vice_results}.

\subsection{Rationale}\label{vice_rationale}

\emph{What drives the connection between metallicity and gas mass?} Analytically, metallicity is not expected to be sensitive to gas mass in itself: for a single zone with constant gas mass, metallicity depends on loading factor as well as inflow and outflow metallicity, as derived in various past works \citep[e.g.][]{dave2012,peng2014,wang2025}. Thus, we require a mechanism which connects gas mass to one of these other factors, or else we must drop the assumption of fixed gas mass.

A rapid increase in metal-poor inflow will lead to rapid decreases in metallicity, and vice-versa \citep[e.g.][]{forbes2014,johnson2020}. However, such fluctuations cannot explain the observed $\mathrm{M_*}$--$\mathrm{M_{gas,in}}$--metallicity relation: gas mass and galaxy size are tightly correlated (Figure \ref{mhismooth}), and galaxy size cannot fluctuate on short timescales. Residual trends between metallicity and dark halo mass \citep{baker2023a,yang2024} likewise discount rapid gas inflow fluctuations as a main cause of metallicity variation, as does the close relationship between inner gas mass and stellar metallicity (Figure \ref{scorrs_no}). A further issue is that massive $\mathrm{\sim 10^{11}M_\odot}$ galaxies are expected to quickly return to equilibrium metallicities after any perturbing event, making the effects of such perturbations less likely to be seen for such galaxies \citep{peng2014}. \citet{wang2025} instead argue inflow metallicities as the key drivers of metallicity variations in massive galaxies, with gas mass assumed to be constant at late times in their analytical models.

On the other hand, if gas mass smoothly varies over time within a star-forming zone, then metallicity is expected to be inversely proportional to the rate of gas mass change, i.e. a rising gas mass results in a lower equilibrium metallicity (and vice-versa). This has been demonstrated \rev{both} analytically \citep[e.g.][]{lilly2013} \rev{and} numerically \citep[e.g.][]{wang2021}. This interpretation in turn implies that gas mass reservoirs are not constant but rather vary on long timescales, a result that has been seen in cosmological simulations \citep{torrey2019}. Thus, if current gas masses broadly correspond to the rate at which gas mass is changing, this could plausibly explain the link between inner gas mass and metallicity. Such an interpretation would imply a close link between the current inner gas mass and the long-term inflow history.

\begin{figure}
\begin{center}
	\includegraphics[trim = 0cm 0cm 0cm 0cm,scale=0.75]{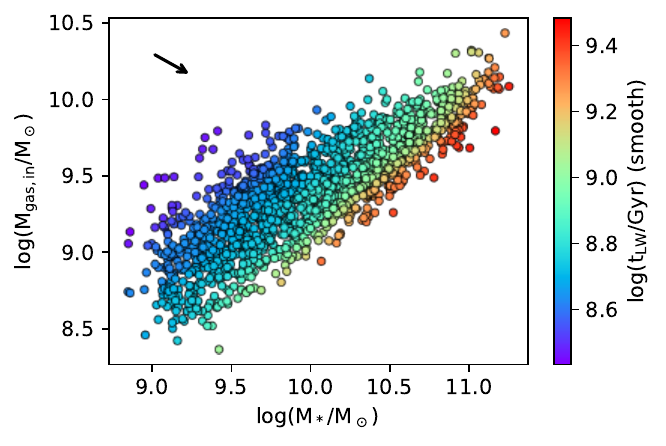} 
	\caption{Inner gas mass vs stellar mass, colored by light-weighted stellar ages at 1~$\mathrm{R_e}$ with smoothing applied. The arrow shows the direction of maximum increase computed from partial correlation coefficients. Lower gas masses are associated with older ages at fixed $\mathrm{M_*}$, supporting a link between inner gas masses and broad star-formation histories.}
	\label{MsMg_age}
	\end{center}
\end{figure}

We plot light-weighted stellar ages at 1~$\mathrm{R_e}$ as a combined function of $\mathrm{M_*}$ and $\mathrm{M_{gas,in}}$ in Figure \ref{MsMg_age}, showing gas-poorer galaxies to be older at fixed stellar mass. Earlier (later) SFHs can in turn be reasonably expected to correspond to earlier (later) inflow histories. Thus, from Figure \ref{MsMg_age}, it is indeed plausible that current inner gas masses are broadly indicative of long-term inflow histories. We therefore argue that lower $\mathrm{M_{gas,in}}$ is associated with earlier average inflow histories at a given $\mathrm{M_*}$, raising the question of whether different inflow histories can explain the connection between $\mathrm{M_{gas,in}}$ and metallicity. This is the primary point that we will investigate through VICE.

Figure \ref{MsMg_age}'s result should not be seen as surprising. \rev{More compact galaxies have been reported to be older at fixed $\mathrm{M_*}$ \citep[e.g.][]{scott2017, barone2020}, while we found in this work a tight correlation between inner gas mass and size (Section \ref{res1}). More compact galaxy disks have also been found to be older at fixed $\mathrm{M_*}$ after bulge--disk decompositions \citep{robotham2022}. These points reiterate} the close connection between $\mathrm{R_e}$ and $\mathrm{M_{gas,in}}$, with a galaxy's star formation history --- and hence long-term inflow history --- relating closely to both its current gas reservoir and size. 

We noted previously that $\mathrm{M_{gas,in}}$ appears to relate even more tightly to N/O than to metallicity (from comparing Figure \ref{scorrs_no}'s top panel to Figure \ref{scorrs}), which long-term inflow history differences can potentially explain due to nitrogen enrichment lagging behind oxygen \citep[e.g.][]{johnson2023}. Thus, we will consider both nitrogen and oxygen in our VICE models.

In summary, we will use VICE to investigate how long-term inflow histories relate to gaseous chemical abundances (O/H, N/O). We will test if differences in long-term inflow can explain the observed close connection between current inner gas masses and chemical abundances. We assume that the one-zone VICE models apply to inner gas masses ($\mathrm{M_{gas,in}}$), meaning that the models apply to the star-forming disk alone. 

\subsection{VICE model construction}\label{vice_generation}

We are aiming here to specifically investigate the role of long-term inflow histories in setting gas abundances, with free parameters kept to a minimum and with inflow histories following a well-motivated functional form. We therefore assumed pristine inflows and constant gas consumption timescales ($\mathrm{t_c = M_{gas,in}/SFR = 4.566\ Gyr}$) throughout our VICE analysis, with our adopted $\mathrm{t_c}$ corresponding to the median value calculated from our sample of galaxies. We then constructed model inflow histories using equation 3 of \citet{wechsler2002}:

\begin{equation}
\mathrm{M(z) = M_0 e^{-p^\prime z}}
\end{equation} \label{weq3}

\noindent where $\mathrm{M_0}$ is the total accreted mass at the present epoch and $\mathrm{p^\prime}$ is an arbitrary power. \citet{wechsler2002} found this formalism to be an excellent descriptor of halo accretion histories in N-body simulations. \rev{We then obtain the inflow rate by differentiating this expression with respect to time\footnote{Performed using the \textit{findiff} \textsc{python} package \citep{findiff}.}.}

Given the above, our VICE models have just three free parameters: the mass loading factor ($\mathrm{\epsilon}$), the total accreted gas mass ($\mathrm{M_0}$) and the inflow history shape (governed by $\mathrm{p^\prime}$ as per Equation 3). For a given $\mathrm{M_0}$, higher $\mathrm{p^\prime}$ corresponds to later inflow histories and higher final gas masses. Thus, we would expect higher $\mathrm{p^\prime}$ to also correspond to younger stellar ages and to larger stellar sizes, with the latter point being due to the observed strong correlation between inner gas mass and $\mathrm{R_e}$.

We ran all VICE models up to 13.3 Gyr, assuming a Kroupa IMF with all stars above 8~$\mathrm{M_\odot}$ exploding as core-collapse supernovae. Following \citet{johnson2020}, who in turn were following
\citet{weinberg2017}, we assumed oxygen yields of 0.015 from CCSNe only. VICE assumes instant recycling for CCSNe, such that the change in oxygen mass from these supernovae is given by $\mathrm{\dot M_O^{CC} = y_{O}^{CC}SFR}$ where $\mathrm{y_{O}^{CC}}$ is the assumed oxygen yield. Following \citet{johnson2021}, we used AGB yields from \citet{cristallo2011} for oxygen, though we note that AGB stars carry little importance for this element \citep[e.g.][]{kobayashi2020}. VICE returns metallicities relative to solar values ($\mathrm{[O/H]}$), which we convert to absolute metallicities using VICE's adopted solar metallicity $\mathrm{12+\log(O/H)_\odot = 8.69}$ \citep{asplund2009}. Our VICE models also track the evolution of nitrogen, with details and results described later in Section \ref{vice_results_n}. 

We first generated four VICE models, given ID numbers of 1-4 for convenience, which we tuned to closely follow the $\mathrm{M_{gas,in}}$--$\mathrm{M_*}$ relation and the mass-metallicity relation. This resulted in the models having progressively lower $\mathrm{p^\prime}$ at higher $\mathrm{M_*}$, consistent with more massive galaxies being older as expected \citep[e.g.][]{gallazzi2005}. We then generated a set of higher--$\mathrm{p^\prime}$ models with identical $\mathrm{\epsilon}$ and with near-identical $\mathrm{M_*}$, representing galaxies with later inflow histories; these models possess higher final gas masses for their $\mathrm{M_*}$. Finally, we generated a set of lower-$\mathrm{p^\prime}$ models which likewise possess the same $\mathrm{\epsilon}$ and similar $\mathrm{M_*}$, representing galaxies with earlier inflow histories; these possess lower gas masses for their $\mathrm{M_*}$. We list the resulting 12 models in Table \ref{vicetable} along with their adopted parameters. We acknowledge that the adopted loading factors (2.6-3.7) are somewhat higher than inferred from simulations \citep[$\sim$unity or below for massive galaxies; e.g.][]{lin2023}; however, the loading factors are highly degenerate with uncertain oxygen yield in our models \citep[e.g.][and refs therein]{johnson2023}, and our absolute metallicities are dependent on the strong-line calibration we adopt, so we caution against interpreting the absolute values of the loading factors in any detail.

\begin{table}
\begin{center}
\begin{tabular}{c c c c c}
Model ID & $\mathrm{\log(M_0)}$ & $\mathrm{\epsilon}$ & $\mathrm{p^\prime}$ & Color\\
\hline
\hline
1 & 11.200 & 2.60 & 0.20 & Red\\ 
2 & 10.850 & 2.80 & 0.32 & Orange\\
3 & 10.500 & 3.15 & 0.50 & Green \\
4 & 10.100 & 3.70 & 0.74 & Blue \\
1 high-$\mathrm{p^\prime}$ & 11.197 & 2.60 & \rev{0.30} & Red\\ 
2 high-$\mathrm{p^\prime}$ & 10.845 & 2.80 & \rev{0.50} & Orange\\
3 high-$\mathrm{p^\prime}$ & 10.492 & 3.15 & \rev{1.00} & Green \\
4 high-$\mathrm{p^\prime}$ & 10.090 & 3.70 & \rev{1.50} & Blue \\
1 low-$\mathrm{p^\prime}$ & 11.204 & 2.60 & \rev{0.10} & Red\\ 
2 low-$\mathrm{p^\prime}$ & 10.857 & 2.80 & \rev{0.15} & Orange\\
3 low-$\mathrm{p^\prime}$ & 10.518 & 3.15 & \rev{0.25} & Green \\
4 low-$\mathrm{p^\prime}$ & 10.126 & 3.70 & \rev{0.40} & Blue \\
\end{tabular}
\end{center}
\caption{Details of VICE chemical evolution models. From left to right: model number, total accreted gas mass at present epoch, power for exponential inflow history, mass loading factor, color of model on corresponding plots.}
\label{vicetable}
\end{table}

\begin{figure*}
\begin{center}
	\includegraphics[trim = 0.2cm 0.5cm 0cm 0cm,scale=0.8]{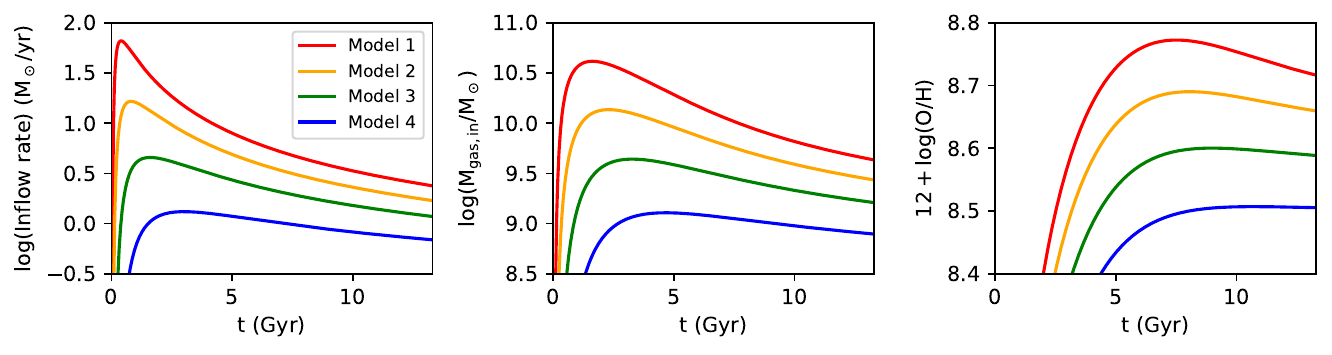} 
	\caption{Evolution of VICE models in terms of inflow rate (left), gas mass (middle) and gas metallicity (right). From models 1-4, $\mathrm{p^\prime}$ and $\epsilon$ increase and the total accreted mass decreases (see Table \ref{vicetable}).}
	\label{vice_models}
	\end{center}
\end{figure*}

We present in Figure \ref{vice_models} the time-evolution of models 1-4 in terms of inflow rate, gas mass and metallicity. Since we assume constant mass-loading factors and constant gas consumption timescales, the time evolution of model metallicities is governed entirely by time evolution in inflow rate. In reality the mass loading factor would be expected to decrease as a galaxy evolved to higher mass, which is not included in our models. This would substantially impact the stellar metallicities (a weighted history of gas metallicity) but is less important for current gaseous abundances, for which the system's current state is the most important consideration. Our purpose for these models is specifically to investigate the impact that inflow variations have on the gaseous chemical abundances.

\begin{figure}
\begin{center}
	\includegraphics[trim = 0.2cm 0.5cm 0cm 0cm,scale=0.8]{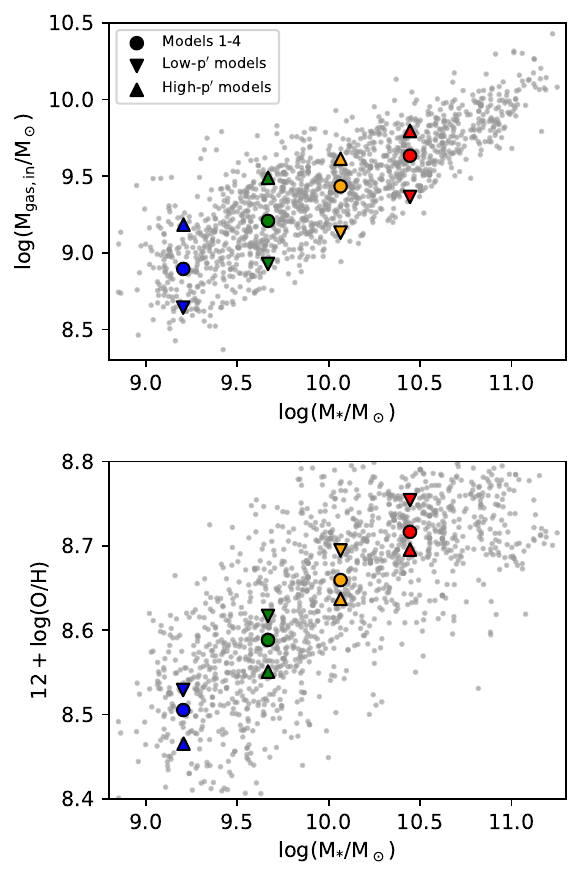} 
	\caption{Final states of the twelve VICE models in terms of inner gas mass vs stellar mass (top) and the mass-metallicity relation (bottom). The MaNGA galaxy sample is shown as small gray points. The models show a good coverage of the data, as well as an inverse trend between gas mass and metallicity at fixed $\mathrm{M_*}$ as indicated by the opposite positions of the triangles in upper and lower plots.}
	\label{vice_scaling}
	\end{center}
\end{figure}

In Figure \ref{vice_scaling}, we compare the final states of the VICE models in terms of stellar mass, inner gas mass and metallicity, to our sample of galaxies.  By construction, the models approximately cover the shape of the observed stellar mass - gas mass, and stellar mass - metallicity relations. The model also recovers the expected inverse trend between gas mass and metallicity at fixed stellar mass. 

\subsection{VICE results}\label{vice_results}

\begin{figure}
\begin{center}
	\includegraphics[trim = 5cm 10cm 0cm 10cm,scale=0.75]{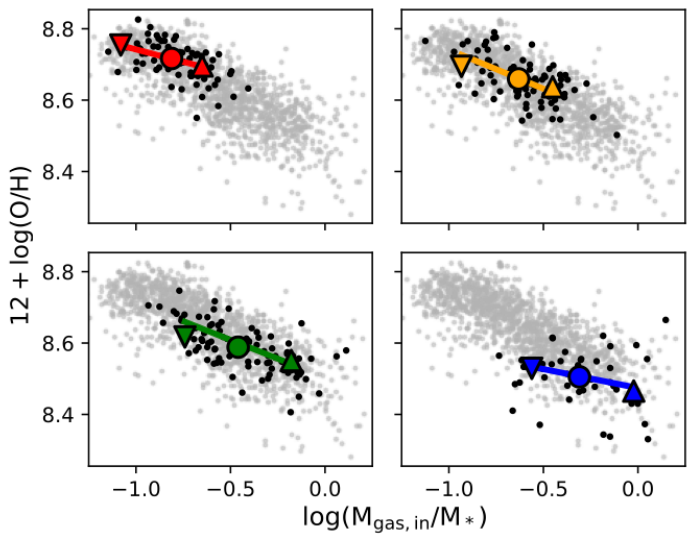} 
	\caption{Final states of the twelve VICE models plotted in terms of inner gas mass vs gas metallicity, with symbols as in Figure \ref{vice_scaling}. The colored lines show straight-line fits to MaNGA galaxies of similar stellar mass (black points), as described in the text. The gray points show the full galaxy sample. }
	\label{vice_gasfrac}
	\end{center}
\end{figure}

We now consider more directly how our VICE models compare with observed galaxies. For models 1-4, we select all MaNGA galaxies with stellar masses within 0.05~dex of that model. We then perform a least absolute deviation\footnote{Performed using sklearn.linear\_model.QuantileRegressor in \textsc{python}, with quantile set to 0.5 and alpha set to 0.} straight-line fit to the $\mathrm{M_{gas,in}/M_*}$--O/H relation for each selected set of galaxies, to trace the connection between gas mass and metallicity at fixed stellar mass. We plot the resulting fits in Figure \ref{vice_gasfrac} along with the twelve models. The models follow the broad trends in a qualitative sense, and they are well within the scatter of the data. 

We show the $\mathrm{M_{gas,in}/M_*}$--O/H slopes fitted from the data in Table \ref{slopetable}, along with showing slopes fitted directly to the VICE models at fixed mass. For the data-derived slopes, we determined errors by taking the standard deviations of 100 Monte Carlo resimulations with bootstrapped residuals. From Table \ref{slopetable}, it is apparent that the intermediate-mass models (models 2 and 3) under-predict the slope of $\mathrm{M_{gas,in}/M_*}$--O/H anti-correlation, while models 1 and 4 obtain good overall agreement with the data. We also note that the data shows significant scatter at low masses especially, which our models cannot account for. 

\begin{table}
\begin{center}
\begin{tabular}{c c c}
Model ID & Model $\mathrm{M_{gas,in}/M_*}$--O/H slope & Data $\mathrm{M_{gas,in}/M_*}$--O/H slope\\
\hline
\hline
1 & $-0.136$ & $-0.138 \pm 0.029$ \\ 
2 & $-0.119$ & $-0.219 \pm 0.023$ \\
3 & $-0.117$ & $-0.210 \pm 0.024$ \\
4 & $-0.118$  & $-0.106 \pm 0.034$ \\
\end{tabular}
\end{center}
\caption{Slopes in the $\mathrm{M_{gas,in}/M_*}$--O/H relation calculates across VICE models at fixed $\mathrm{M_*}$ and across corresponding MaNGA subsamples.}
\label{slopetable}
\end{table}

Data--model differences are unsurprising given our models' simplicity. In particular, gas inflows into more massive galaxies are likely not pristine \citep[e.g.][]{belfiore2016,sm2019,bassini2024}, and gaseous outflows also appear from observations to be enriched compared to the ISM \citep{chisholm2018}. Galaxies also show non-negligible scatter in gas consumption timescales at fixed $\mathrm{M_*}$ \cite{saintonge2017}, which may lead to increased chemical abundance scatter. \rev{An additional caveat is that our models' assumption of a direct SFR--$\mathrm{M_{gas,in}}$ scaling implicitly predicts a positive size--SFR correlation at fixed $\mathrm{M_*}$,  which is also predicted in cosmological simulations \citep{sm2018a,ma2024} but which is not actually supported by data for star-forming galaxies \citep{wuyts2011,boardman2025}}. Nonetheless, it is apparent from our models that broad inflow variations can contribute significantly to metallicity variations. In turn, broad inflow variations could play a key role in observed $\mathrm{M_{gas,in}/M_*}$--O/H trends. 

\subsubsection{Effective yields}

As an interesting aside, we consider our models' predictions for effective yields in comparison to data. Effective yields are derived via the simple closed-box formalism, in which a galaxy is modeled as a single star-forming zone with neither inflow nor outflow, instantaneous mixing and instantaneous recycling \citep[e.g.][]{pagel1975,edmunds1990}. This model is known to be inaccurate for the Milky Way, due to it over-predicting the proportion of metal-poor stars \citep[e.g.][]{vdb1962,pagel1975}, but it is potentially more accurate for star-forming galaxy populations more generally \citep{valeasari2009,greener2021}. In closed box models there is an analytical relation between the metal mass fraction ($\mathrm{Z}$) and the gas-to-total mass ratio $\mathrm{\mu}$, where $\mathrm{\mu = M_{gas}/(M_* + M_{gas})}$:

\begin{equation}
    \mathrm{Z = y/\ln(\mu^{-1})}
\end{equation}

\noindent where $\mathrm{y}$ describes the physical metal yield and $\mathrm{Z}$ the metal mass fraction. An `effective' yield ($\mathrm{y_{eff}}$) can thus be determined observationally from the closed box formalism as

\begin{equation}
    \mathrm{y_{eff} = Z\ln(\mu^{-1}) = Z\ln\left(1 + \frac{M_*}{M_{gas}}\right)}
\end{equation}\label{yeff}

\noindent where we convert gas metallicities to metal mass fractions by assuming a solar abundance mixture with $\mathrm{Z_\odot}$ = 0.014 \citep{asplund2009}. 

In the case of no inflow or outflow, $\mathrm{y_{eff}}$ will be equal to the physical metal yield $\mathrm{y}$. More generally, we would expect $\mathrm{y_{eff} < y}$, with the lower $\mathrm{y_{eff}}$ reflecting the impact of inflow and/or outflow on a galaxy's chemical evolution \citep[e.g.][]{edmunds1990}. Variations in $\mathrm{y_{eff}}$ with other galaxy parameters can indicate varying relative impacts of inflow and outflow, motivating the study of $\mathrm{y_{eff}}$ scaling relations. $\mathrm{y_{eff}}$ was reported to scale positively with baryonic mass in \citet{tremonti2004}, who used SFRs as gas mass proxies in nearby galaxies. Tests with direct gas tracers instead show a turnover mass at $\mathrm{\sim10^{10}M_\odot}$, beyond which effective yield declines with baryonic mass \citep{ll2019}; this is similar to what is observed at higher redshifts \citep{erb2006,mannucci2009} and also in simulations \citep{ll2019,zerbo2024} .  

\begin{figure}
\begin{center}
	\includegraphics[trim = 0.2cm 0.5cm 0cm 0cm,scale=0.8]{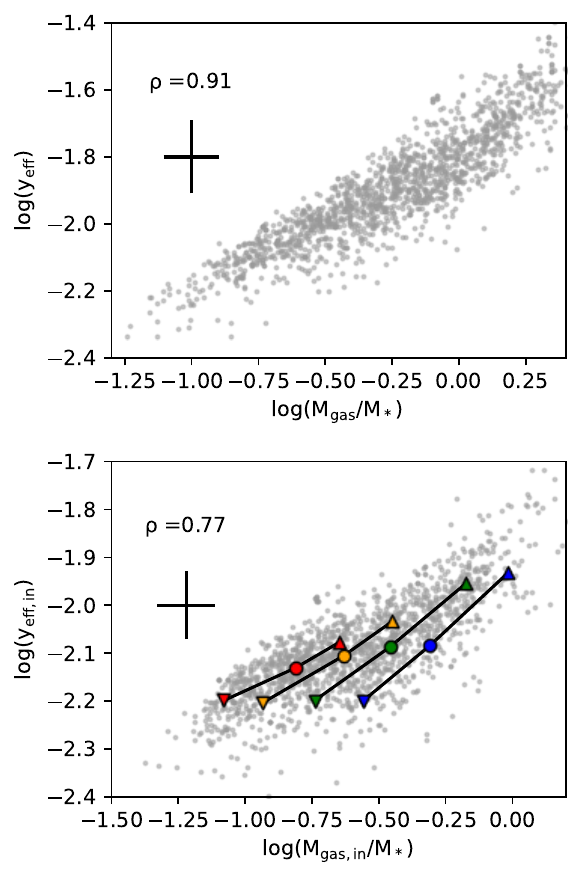} 
	\caption{Top: effective yield vs gas fraction, plotted for our MaNGA sample. Bottom: inner effective yield vs inner gas fraction, plotted for the data and for the twelve VICE models with symbols as in Figure \ref{vice_scaling}. Error bars show the median data errors, with the Spearman correlation coefficient also displayed for the data.}
	\label{vice_yeff}
	\end{center}
\end{figure}

To facilitate comparisons between our models and the MaNGA data, we define an `inner' effective yield $\mathrm{y_{eff.in}}$ in which $\mathrm{M_{gas}}$ is replaced with $\mathrm{M_{gas,in}}$ in Equation 5. We plot $\mathrm{y_{eff}}$ and $\mathrm{y_{eff,in}}$ against their corresponding gas-to-stellar mass ratios in Figure \ref{vice_yeff}, with the VICE models also shown in the latter case. We find both effective yield measures to vary tightly with the gas fractions, which in the case of $\mathrm{y_{eff,in}}$ is well-reproduced by the VICE models.

We also find that both effective yield measures show non-negligible anti-correlations with $\mathrm{M_*}$, suggesting that lower-mass galaxies follow closed-box expectations more closely. Taking Spearman correlations, we find $\mathrm{\rho = -0.51}$ between $\mathrm{y_{eff}}$ and $\mathrm{M_*}$ and we find $\mathrm{\rho = -0.43}$ between $\mathrm{y_{eff,in}}$ and $\mathrm{M_*}$. This is consistent with metallicity histories derived from CALIFA\footnote{Calar Alto Legacy Integral Field Area Survey} \citet{mn2020} and MaNGA \citet{greener2021} data, which are more similar to closed-box expectations for lower-mass galaxies, and it is also consistent with the finding of \citet{frasermckelvie2022} that gas--stellar metallicity offsets decline with increasing stellar mass.

Observationally, our effective yield results provide further evidence that gas-rich and gas-poor galaxies experience fundamentally different chemical evolution histories. Our VICE models, meanwhile, suggest that broadly different inflow histories can explain the observed trends.

\subsubsection{Nitrogen}\label{vice_results_n}

For nitrogen, we used the default SNIa yield of $\mathrm{6.43\times 10^{-9}}$, effectively meaning that SNIa contribute negligibly to nitrogen enrichment. We also assumed core collapse supernovae to possess nitrogen yields independent of oxygen \citep[e.g.][]{vincenzo2016,johnson2023}, meaning that the nitrogen yield from core collapse supernovae can be described by a single yield parameter $\mathrm{y_{N}^{CC}}$. For AGB stars meanwhile, we used the following functional form:

\begin{equation}
\mathrm{Y_N^{AGB}= \xi \left(\frac{M_\star }{M_\odot}\right)\left(\frac{Z_\star}{Z_\odot}\right)^\beta}
\end{equation} 

\noindent where $\mathrm{M_\star}$ and $\mathrm{Z_\star}$ indicate the masses and metal mass fractions of individual stars. $\mathrm{\xi}$ is an arbitrary scale factor and $\mathrm{\beta}$ an arbitrary power to set the metallicity-dependence of the nitrogen yield, with Equation 6 corresponding to equation 3 of \citet{johnson2023} when $\mathrm{\beta = 1}$. The mass yield is given by $\mathrm{M_\star Y_N^{AGB}}$, with the IMF-averaged yield $\mathrm{y_N^{AGB}}$ then found by integrating over masses of 1--8 $\mathrm{M_\odot}$. We first use the fiducial CCSNe and AGB star yields presented in \citet{johnson2023}, wherein $\mathrm{y_{N}^{CC} = 3.6\times 10^{-4}}$, $\mathrm{\xi = 9\times10^{-4}}$ and $\mathrm{\beta} = 1$. We refer to these as the `J23 yields'.  We converted from VICE's output $\mathrm{[N/O]}$ using the solar abundance $\mathrm{12+\log(N/H)_\odot = 7.83}$ \citep{asplund2009}. 

\begin{figure}
\begin{center}
	\includegraphics[trim = 0.2cm 0.5cm 0cm 0cm,scale=0.8]{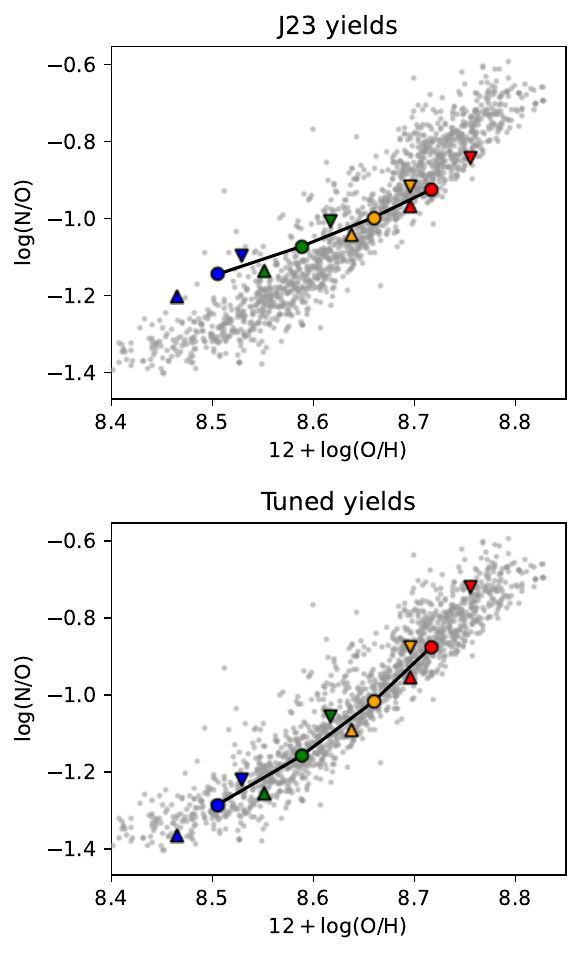} 
	\caption{N/O--metallicity relation for our twelve models with the J23 yields (top) and tuned yields (bottom) for our twelve VICE models and for the MaNGA sample. Symbols are as in Figure \ref{vice_scaling}. To help guide the eye, a black line is plotted between models 1-4. We find that long-term inflow history differences cause N/O to vary at fixed O/H, in line with analytical expectations.}
	\label{vice_nooh}
	\end{center}
\end{figure}

We plot our twelve VICE models in N/O--metallicity space with the J23 yields in the top panel of Figure \ref{vice_nooh}, with the MaNGA data also displayed. Our VICE models demonstrate that differences in inflow histories induce N/O variations \emph{at fixed metallicity}. This is expected, due to nitrogen enrichment lagging behind oxygen. As a reminder, N/O varies more tightly with $\mathrm{M_*}$ and $\mathrm{M_{gas,in}}$ than does metallicity (from comparing the top panel of Figure \ref{scorrs_no} to Figure \ref{scorrs}). Figure \ref{vice_nooh} demonstrates that long-term inflow history differences can explain this: higher (lower) $\mathrm{p^\prime}$ models are shifted to lower (higher) N/O for their gas metallicity, as opposed to models shifting along the direction of the N/O--metallcity relation. This also offers an explanation for the results of \citet{boardman2024a}, wherein $\mathrm{\Phi_e}$ was more strongly correlated with N/O than with O/H. 

The modelled N/O--metallicity relation is much flatter than the data when using the J23 yields, which is to be expected: \citet{johnson2023} benchmarked their fiducial model to the \citet{Dopita_2016_EmLineDiagnostic} N/O--metallicity relation, which is significantly flatter than our observed relation. It should be noted that N/O and metallicity are highly calibrator-dependent when using strong line methods \citep[e.g.][]{kewley2008,scudder2021,florido2022}, with strong line methods also artificially reducing the scatter in the N/O--metallicity relation \citep{valeasari2016}. A full investigation of this point would require the use of weaker emission lines, such that N/O and metallicity could be determined either from the direct method or from suitable photoionisation models \citep[e.g.][]{valeasari2016}. Using such lines in MaNGA would require the use of stacked spectra, which is beyond the scope of this present work. However, it remains worthwhile to consider how our VICE models behave in the case where the N/O--metallicity relation is recovered.

Given the above, we adopted a second set of nitrogen yields tuned to match the data: we set $\mathrm{y_{N}^{CC} = 3\times 10^{-4}}$, $\mathrm{\xi = 1 \times10^{-3}}$ and $\mathrm{\beta = 2}$, steepening the predicted N/O--metallicity relation. We refer to the resulting yields as the `tuned yields'. We show the results in the bottom panel of Figure \ref{vice_nooh}: inflow history differences again cause N/O to rise with fixed metallicity as expected, with the models remaining within the data's scatter. Thus, we argue long-term inflow history differences to be a likely explanation for why N/O scaling relations appear so tight.

\section{Discussion}\label{disc}

We combined MaNGA integral-field spectroscopy with HI-MaNGA radio follow-up to investigate the connection between galaxies' gas metallicities and gas masses. We also investigated gas masses and total dynamical masses across the stellar mass-size plane, motivated by the striking reported correlation between gas metallicity and stellar concentration \citep[$\mathrm{\Phi_e = M_*/R_e}$; e.g.][]{deugenio2018,sm2024}. Our key results from this work are as follows:

\begin{itemize}
\item Central dynamical masses vary across the mass-size plane, such that more extended galaxies possess higher dynamical masses within 1~$\mathrm{R_e}$ (Figure \ref{mre_dyn}). This means that $\Phi_e$ is \emph{not} a good proxy for the depth of galaxies' gravitational potential wells, contrary to what is sometimes assumed. $\Phi_e$ simply represents the concentration of stellar mass, and we suggest the phrase `stellar concentration' to describe it.
\item Galaxies' total HI masses vary as functions of both $\mathrm{M_*}$ and $\mathrm{R_e}$, such that more extended galaxies possess higher gas masses at fixed stellar mass (Figure \ref{mhismooth}, top panel). This is consistent with the findings of \citet{pan2021}.
\item As a result of the above, galaxies' inner HI masses ($\mathrm{M_{HI,in}}$, within the 90\% light radius) are tightly related to galaxies' size (Figure \ref{mhismooth}, bottom panel). Extended galaxies possess much more gas within their optical extents than compact galaxies.
\item After accounting for stellar mass, gas metallicity relates most closely to the inner (atomic plus molecular) gas mass; we showed this using correlation coefficients (Figure \ref{scorrs}) and with a random forest analysis (Figure \ref{randomforest_main}). 
\item We obtain equivalent results for gas-phase N/O and for stellar metallicity (Figure \ref{scorrs_no}), with N/O actually producing stronger correlations than gas metallicity. This suggests that $\mathrm{M_{gas,in}}$ is sensitive to a galaxy's longer term chemical evolution history, in addition to being highly indicative of a galaxy's current chemical state.
\item $\Phi_e$ is tightly connected to a galaxy's position in $\mathrm{M_*-M_{gas,in}}$ space (Figure \ref{inmgas_phi_pcc}). This appears to be the true reason for the $\Phi_e$--metallicity relation.
\item We used one-zone chemical evolution models to show how differences in long term gas inflow histories can produce anti-correlations between gas mass and metallicity (Figure \ref{vice_gasfrac}) along with even stronger anti-correlations between gas mass and N/O (Figure \ref{vice_nooh}). 
\end{itemize}

A potential concern with this work is that $\mathrm{M_{HI,in}}$ is connected to galaxies' sizes by our adopted method, leaving $\mathrm{M_{gas,in}}$ connected to size by extension. Thus, one might be tempted to interpret the $\mathrm{M_{gas,in}}$--metallicity connection as being driven by an underlying size--metallicity connection. We reiterate that, at fixed $\mathrm{M_*}$, the metallicity is predicted more effectively by $\mathrm{M_{gas,in}}$ and $\mathrm{M_{HI,in}}$ than by size, making it highly unlikely that the predictive strength of $\mathrm{M_{gas,in}}$ comes from size alone. It is also intuitive that gas mass shows a stronger connection to metallicity when both quantities are calculated over similar physical scales, as argued by \citet{chen2022a} for $\mathrm{M_{HI,in}}$, with the outskirts of galaxies containing substantial HI gas that is not relevant to star-formation \citep[e.g.][]{wang2014}.

We present a physical interpretation of our findings in Section \ref{interpretation}. We then compare with previous metallicity results in Section \ref{otherworks}, before discussing the key implications for future works in Section \ref{implications}.

\subsection{Interpretation}\label{interpretation}

\subsubsection{Galaxy metallicities and gas reservoirs}

Our results have unified two largely separate branches of galaxy chemical abundance results: those relating to the size-metallicity connection at fixed $\mathrm{M_*}$ \citep[e.g.][]{ellison2008,deugenio2018,vaughan2022,sm2024,boardman2024a} and those on how metallicity connects to SFR or gas mass at fixed $\mathrm{M_*}$ \citep[e.g.][]{ellison2008,mannucci2010,ll2010,bothwell2013,bothwell2016}. In particular, our results imply that the tight $\Phi_e$--metallicity relation \citep[e.g.][]{deugenio2018} is simply a projection of a more fundamental relation between stellar mass, inner gas mass and gas metallicity. 

Our results demonstrate that gas masses of massive ($\mathrm{>10^{8.8}M_\odot}$) galaxies cannot experience significant rapid fluctuations in the nearby Universe: galaxies display a close connection between their gas masses and optical sizes, and sizes cannot fluctuate on short timescales. This point is entirely consistent with past literature: similar arguments can be made from the residual anticorrelation between SFR and stellar metallicity \citep{looser2024}, from residual trends between metallicity and dark halo mass \citep{baker2023a,yang2024}, and from the expected timescales of SFR variations in simulations \citep{tachella2016}\footnote{The cosmological simulations of \citet{tachella2016} suggest SFR variations on timescales of 0.4 Hubble times, which at z = 0 corresponds to $\sim$5 Gyr.}. 

Using VICE one-zone chemical evolution models, we were able to qualitatively reproduce trends in our sample involving $\mathrm{M_*}$, $\mathrm{M_{gas,in}}$, gas metallicity and N/O. We did this by invoking long-term differences in inflow histories at fixed $\mathrm{M_*}$, wherein gas-richer galaxies experience later inflow histories. Given that inner gas mass correlates closely to galaxy size (Section \ref{res1}), this also offers an explanation as to why more compact star-forming galaxies are more metal-rich at fixed $\mathrm{M_*}$. 

In a quantitative sense, our VICE models do not entirely reproduce the observed metallicity behavior (Table \ref{slopetable}): our intermediate-mass models under-predict the slope of the anti-correlation between gas mass and metallicity at fixed $\mathrm{M_*}$, with the data showing non-negligible scatter (especially at low stellar masses) which the models cannot account for. At lower stellar masses this could simply be due to the range of individual galaxy inflow histories increasing the metallicity scatter, with scatter in gas consumption timescales being another possible factor. For higher-mass galaxies however, an additional explanation is clearly called for. 

A key limitation of our models is that we assume pristine inflows --- an assumption which is likely inaccurate for real massive galaxies. While lower-metallicity galaxies display spatially local anticorrelations between SFR and metallicity \citep{sm2019,bulichi2023}, suggesting infall by metal-poor gas, high-metallicity galaxies instead have positive local SFR--metallicity correlations \citep{sm2019}. As argued by \citet{sm2019}, this suggests that high-metallicity galaxies receive inflows of recycled, enriched material. Inflow metallicities also appear to be the driving factor behind the size-metallicity connection in \textsc{l-galaxies} semi-analytic models \citep{ayromlou2021}, as demonstrated by \citet{wang2025}. Therefore, a combination of long-term inflow history differences \textit{and} inflow metallicity differences are likely relevant to fully understanding chemical abundances in massive galaxies.

\subsubsection{Galaxy-halo connection}

The dark halo properties of compact and extended galaxies are also relevant to consider, with our finding that extended galaxies contain significantly more total mass within their central effective radii (Section \ref{res2}). Weak lensing measurements also suggest more extended \rev{star-forming} galaxies to possess more dark matter \citep{charlton2017}, in good consistency with our observational results \footnote{We note that \rev{galaxy clustering} instead \rev{suggests} gas mass to negatively correlate with halo mass \rev{for star-forming galaxies up to $\mathrm{10^{10.5}\ M_\odot}$} \citep{yang2024}. This discrepancy may be due to
differences in halo concentrations \citep{charlton2017,taylor2020},
however, understanding this result is beyond the scope of this discussion.}, \rev{though the opposite trend is seen when viewing galaxies as a single population \citep{taylor2020}}.

Our results are also in good consistency with \textsc{l-galaxies} semi-analytic model predictions \citep{wang2025}. Compact galaxies in \textsc{l-galaxies} form stars more efficiently and so reach higher $\mathrm{M_*}$ at fixed halo mass, which is equivalent to them having lower halo mass at fixed $\mathrm{M_*}$. If this is the case, then we expect extended galaxies to more effectively accrete metal-poor gas from their surroundings, producing more massive gas reservoirs along with lower metallicities. Compact galaxies, by contrast, would accrete less metal-poor gas from their surroundings due to them possessing less dark matter within their star-forming disks; this would produce lower gas reservoir masses along with higher metallicities. We caution however that, while this is consistent with semi-analytic models \citep{wang2025}, it is potentially at odds with hydrodynamical simulations: \citet{vdv2011} for instance show gas accretion onto galaxies to decline with halo mass beyond halo masses of $\mathrm{\sim 10^{12.5}M_\odot}$.  

If compact galaxies indeed accrete less metal-poor gas, then an additional consequence is that compact galaxies' inflows will contain a larger proportion of recycled gas, whereas extended galaxies' inflows will instead consist largely of metal-poorer ex-situ gas. This would immediately create a link between gas masses, sizes and metallicities at fixed stellar mass. 

\subsubsection{Unified physical interpretation}

To summarize our results and discussion thus far: within their optical extents at fixed $\mathrm{M_*}$, extended galaxies possess lower gaseous and stellar metallicities, lower gaseous N/O ratios, more gas and more dark matter within their optical extents from observations. In turn, compact galaxies at fixed $\mathrm{M_*}$ possess higher gaseous and stellar metallicities, higher N/O ratios, less HI gas, and less dark matter within their optical extents. From chemical evolution models, we argued that our results on chemical abundances and gas mass can be largely attributed to differences in long-term inflow histories: more compact galaxies possess earlier SFHs and thus earlier inflow histories, resulting in more quickly-declining reservoirs and thus producing lower current gas masses and higher metallicities. Results from the literature also suggest that increased inflow metallicities could be relevant for higher stellar mass galaxies. 

\begin{figure*}
\begin{center}
	\includegraphics[trim = 0.2cm 0.5cm 0cm 0cm,scale=0.5]{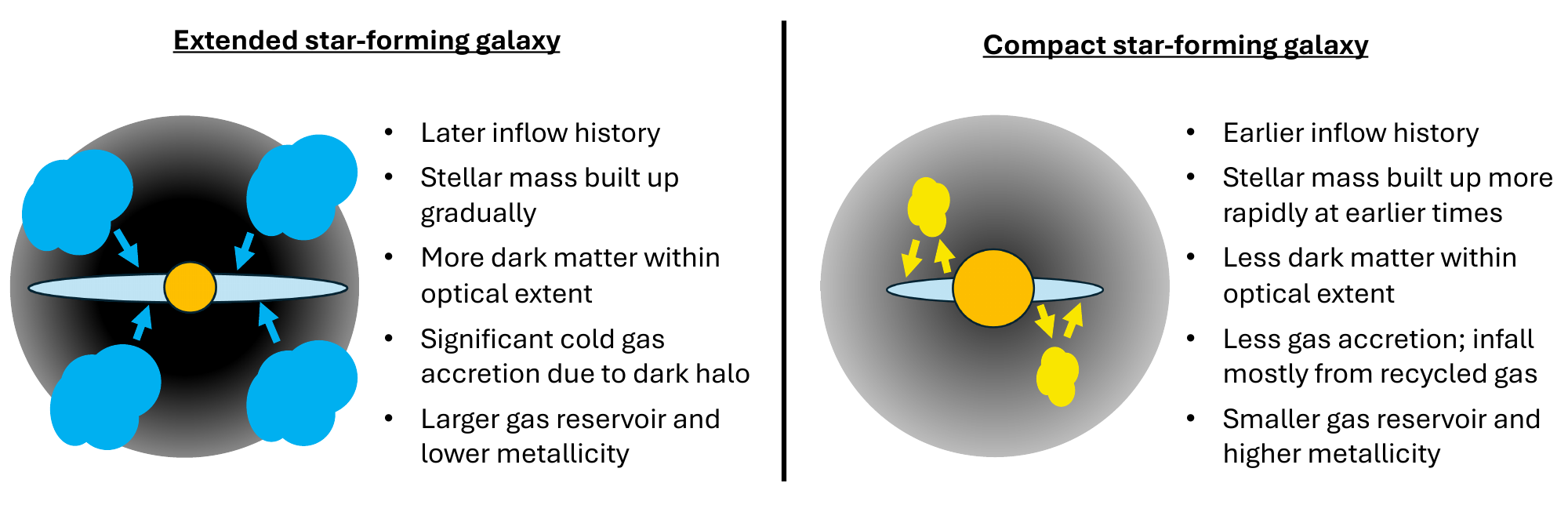} 
	\caption{The physical interpretation of our results for the  chemical evolution of extended and compact star-forming galaxies. Extended galaxies contain more dark matter within their optical extents, resulting in steady late-time accretion of cold metal-poor gas (blue clouds); this results in more massive gas reservoirs and lower present-day metallicities. Compact star-forming galaxies possess less dark matter within their optical extents, resulting in steadily reducing accretion of gas over time, producing less massive gas reservoirs and higher metallicities. Compact galaxies may also mostly receive inflows from enriched recycled gas (yellow clouds) further increasing their metallicity.  In both panels, the black color scale represents the density of dark matter within galaxies' optical extents and the galaxy pictorials reflect the fact that compact star-forming galaxies possess higher bulge-to-total ratios \citep{boardman2023}.}
	\label{stellarconc_cartoon}
	\end{center}
\end{figure*}

In Figure \ref{stellarconc_cartoon}, we present a physical scenario to explain our findings. At any given mass, more extended galaxies possess more dark matter within their star-forming regions, resulting in them maintaining a steady inflow of metal-poor gas to the present day (Figure \ref{vice_models}). Compact galaxies, by contrast, are receiving an ever-diminishing supply of gas. Both the rate of decline in gas inflows, as well as the pre-enrichment of the gas that is inflowing, lead to higher gas metallicities than for extended galaxies. 

\subsection{Comparison with other works}\label{otherworks}

Our metallicity findings build upon \citet{chen2022a}, who first showed the $\mathrm{M_*}$--$\mathrm{M_{HI,in}}$--metallicity relation for star-forming MaNGA galaxies. We demonstrated this relation to be connected to the tight three-way relationship between stellar mass, size and gas metallicity --- a relation itself reported in numerous studies \citep[e.g.][]{ellison2008,deugenio2018,sm2024}. Our results suggest the size-metallicity connection to be a projection of a more fundamental connection between metallicity and gas mass, similar to what has been argued previously for the FMR \citep[e.g.][]{bothwell2013}. We suggest that previous results of a connection between size and gas metallicity, along with results of a connection between size and stellar metallicity \citep[e.g.][]{scott2017,barone2018,barone2020,vaughan2022,looser2024}, should be reinterpreted in light of our findings

To the best of our knowledge, this is the first work showing an explicit link at fixed stellar mass between galaxies' HI masses and their effective radii. However, this result follows naturally from \citet{pan2021}, who obtain an equivalent result using isophotal galaxy sizes\footnote{Defined in \citet{pan2021} as the size at which the r-band surface brightness reaches 25~$\mathrm{mag/arcsec^2}$.}. Such a result also follows logically from other previous literature, which has shown HI mass to correlate with various optical size measures \citep[e.g.][]{chamba2024} while also displaying links with galaxy parameters including stellar mass, global stellar density, optical color and optical concentration \citep[e.g.][]{bothun1984,kannappan2004,wu2020,li2022}. In addition, larger optical sizes are associated with larger average gas disks \citep[e.g.][]{broeils1997,wilman2020}, with the size of HI disks known to correlate tightly with HI masses \citep[e.g.][]{broeils1997,swaters2002,wang2020,rajohnson2022}. Thus, the existence of a strong $\mathrm{M_{HI}}$--$\mathrm{R_e}$ connection should not be viewed as surprising.

At first glance, our results may seem inconsistent with \citet{sm2024}, who perform an RF analysis on MaNGA data and find $\Phi_e$ to have a high importance for determining gas metallicity. This difference is entirely due to the inclusion of inner gas masses in our analysis. As shown in Appendix ~\ref{rfreduced}, simply excluding $\mathrm{M_{gas,in}^{0.7}/M_*}$ and $\mathrm{M_{HI,in}^{0.6}/M_*}$ as RF features results in $\Phi_e$ obtaining the highest importance, which can be understood as being due to its tight association with a galaxy's position in $\mathrm{M_{gas,in}}$--$\mathrm{M_*}$ space (Figure \ref{inmgas_phi_pcc}. 

A minor inconsistency does exist between our work and \citet{sm2024}, in that we find $\mathrm{M_*/R_e^{0.9}}$ to best capture the size-dependence on metallicity. We found this using Spearman correlation coefficients, as shown in Figure \ref{scorrs}. \citet{sm2024} instead find $\mathrm{M_*/R_e^{0.6}}$ to be a more constraining parameter, from an expanded RF analysis of their MaNGA sample. Sample selection and method differences may both be factors here. We note that we prefer using correlation coefficients for this kind of test, due to the potential for overfitting in the RF method. 

Our results also initially appear in tension with \citet{baker2023a}, who find stellar mass to dominate over $\Phi_e$ when using a random forest to predict gas metallicities. This apparent tension is due to the different metallicity measures employed: we specifically use the metallicity \textit{at} 1 $\mathrm{R_e}$ while \citet{baker2023a} use a weighted average \textit{within} $\mathrm{1 R_e}$. As shown in \citet{koller2025}, inner MaNGA gas metallicities trend mostly with $M_*$ while MaNGA metallicities at $\sim$1~$\mathrm{R_e}$ and beyond trend mostly with $\Phi_e$. Given the close correlation between 1~$\mathrm{R_e}$ galaxy properties and global galaxy properties \citep{gonzalezdelgado2014,gonzalezdelgado2015,sanchez2016b}, we argue that 1~$\mathrm{R_e}$ chemical abundances are more suitable for the kind of analysis presented in this work. SDSS galaxies also display tight $\Phi_e$--metallicity correlations when aperture effects are accounted for, as shown in \citet{deugenio2018}.

The idea that gas reservoirs vary over long $(\mathrm{>\sim}$~Gyr) timescales at late times --- a key assumption of our VICE models --- has some support from observational works. Observations have repeatedly shown the star-forming sequence to possess higher normalizations (i.e. higher sSFRS) at earlier lookback times \citep[e.g.][and references therein]{wuyts2011,madau2014}. Recent findings have suggested similar behavior in observed HI reservoirs, with HI masses observed to decline in galaxy populations from $\mathrm{z \sim 1}$ until today \citep[e.g.][]{chowdhury2022,bianchetti2025,depalma2025}.

Our dynamical mass results agree well with the predictions of \citet{wang2025}, who use MaNGA and semi-analytic models to probe the connection between size and stellar metallicity at fixed $\mathrm{M_*}$. They find that, at $\mathrm{M_* \leq 10^{10.5}M_\odot}$, the connection between size and stellar metallicity is best-understood as arising from star-formation efficiency (SFE) differences: the higher SFEs of compact galaxies leads to accelerated early enrichment and locks more metals up into stars at early times, while also producing lower dark halo masses at fixed $\mathrm{M_*}$. This offers a further explanation for our results concerning stellar metallicity, for which we found a close connection to inner gas mass (and by extension to galaxy sizes). However, this explanation is much less relevant for gaseous abundances, which specifically represent the current state of gas reservoirs. 

For higher stellar masses, our interpretation is similar to \citet{wang2025} but is not identical. For $\mathrm{M_* > 10^{10.5}M_\odot}$, \citet{wang2025} argue both stellar and gaseous metallicities to be driven by the metallicity of inflowing gas. We instead argue the main metallicity driver to be the gas reservoir histories, albeit with inflow metallicity potentially being a connected secondary factor.  These different interpretations arise due to different assumptions in our respective works: \citet{wang2025} did not consider variations in gas inflow rate in their gas regulator model, whereas we assume that gas masses smoothly vary over time. We reiterate that galaxy sSFRs have been found to be higher at higher redshifts \citep[e.g.][]{wuyts2011}, supporting the possibility of more massive gas reservoirs in galaxies' pasts. We also reiterate that gas-richer galaxies are younger at fixed $\mathrm{M_*}$ (Figure \ref{MsMg_age}), supporting the idea that inner gas masses connect to long-term SFHs and by extension to long-term inflow histories.

\subsection{Implications for past and future work}\label{implications}

The $\mathrm{\Phi_e}$ parameter ($\mathrm{M_*/R_e}$) has attracted much attention in recent metallicity studies, due to a tight observed correlation with gaseous and stellar metallicity \citep{barone2018,deugenio2018,barone2020,vaughan2022,sm2024} and gaseous N/O \citep{boardman2024a}. In the past, this parameter has been interpreted as a proxy for gravitational potential depth \citep[e.g.][]{deugenio2018,vaughan2022,sm2024}. However, our results show that $\Phi_e$ should not be interpreted this way: at fixed stellar mass, more extended galaxies possess higher central dynamical masses (Figure \ref{mre_dyn}) and thus are expected to have stronger potentials. 

We instead argue that $\Phi_e$ is best interpreted as representing the concentration of a galaxy's stellar mass, and we therefore suggest the phrase `stellar concentration' to describe it. This parameter was argued to be a \rev{powerful} proxy for SFH in \citet{boardman2025}, based on its correlations with stellar and gaseous chemical abundances combined with its correlation with stellar age,  which would make it a close proxy of galaxies' gas inflow histories by extension. This connection with gas inflow histories, we argue, is the true reason for the tight connection between $\Phi_e$ and chemical abundances.

In light of our findings, various observational tests would be fruitful to carry out. It would be particularly useful to check our findings with spatially-resolved HI mass measurements, which for instance could be done with the WEAVE-Apertif survey \citep{jin2024} and WALLABY \citep{koribalski2020,reynolds2023}. It would likewise be useful to investigate how the connection between metallicity and HI mass (both inner and total) evolves over cosmic time, by utilising stacking with current instruments \citep[e.g.][]{sinigaglia2024,depalma2025} along with leveraging future facilities such as the Square Kilometre Array \citep[SKA;][]{braun2015}. It will be important to investigate scaling relations between HI gas mass and photometric properties at higher redshifts \citep[e.g.][]{chowdhury2024}; this will allow for more precise interpretation of galaxy metallicities in cases where gas masses cannot be measured, since photometric properties are comparably easy to determine. Finally, as in \citet{boardman2025}, we advocate studying the size--metallicity connection at higher redshifts; this in our view may be more fruitful in isolation than studying the FMR, given the greater importance of stellar concentration over SFR for interpreting the scatter of the MZR \citep{ma2024a,boardman2025}.

The observational results in this work also provide stringent tests for simulations such as EAGLE \citep{schaye2015}, IllustrisTNG \citep{tng1,tng2,tng3,tng4,tng5,tng6,tng7}, SIMBA \citep{dave2019,hough2023} and COLIBRE \citep{schaye2025}. Of particular note here is the behavior of inner dark halo masses across the stellar mass--size plane, along with the close three-way connection between metallicity, gas mass and size at fixed stellar mass. Should these tests be passed, then simulations could also help distinguish between the scenarios of \citet{wang2025} and this present work. Simulations would allow one to test if late-time gas masses correspond broadly to long-term inflow histories (or equivalently to gas mass histories), which is a critical prediction of our proposed scenario.

\section{Summary and Conclusion}\label{summary}

Galaxy chemical abundances are key tracers of galaxy evolution, connecting directly to galaxies' formation histories. Here, we used MaNGA spectroscopy and HI-MaNGA radio followup to investigate how chemical abundances relate to a range of galaxy parameters.  We also examined HI masses and dynamical masses across the stellar mass-size plane, to better understand how this plane should be interpreted. 

Our analysis mostly focused on gas-phase metallicity, $\mathrm{12 + \log(O/H)}$, though we also considered gaseous N/O along with light-weighted stellar metallicity. All of our abundance values were calculated at 1 effective radius, which has been shown previously to be a good approximation for global quantities which is also resistant to aperture effects. We used Spearman correlation coefficients to analyse how chemical abundances relate to stellar mass, half-light radius, global HI mass and inner (within the 90\% light radius) HI mass, and the total (atomic plus molecular) inner gas mass. We also employed a random forest analysis to determine which galaxy parameters are most intrinsically related to metallicity and abundances.

From both correlation coefficients and a random forest analysis, we found gas metallicity at fixed stellar mass to relate most fundamentally to the total inner gas mass. We obtained equivalent results for N/O and stellar metallicity, suggesting that inner gas masses are indicative of galaxies' long-term chemical evolution histories. We also pointed out a close connection between galaxies' sizes and their inner HI masses; this means that previously-reported size--metallicity anticorrelations \citep[e.g.][]{ellison2008,deugenio2018} should be interpreted in terms of gas reservoirs. In addition, we found that more extended galaxies possess higher central dynamical masses at fixed stellar mass; this is consistent with weak lensing measurements, and it suggests more extended star-forming galaxies to possess more dark matter within their optical extents. 

We interpret the connection between size and gas mass at fixed stellar mass as arising from differences in long-term gas inflow histories: more compact star-forming galaxies accrete more gas mass earlier, leading to their higher stellar ages \citep{barone2020}, and resulting in lower gas masses at late times. We use a series of VICE chemical evolution models to show that the same differences in long-term gas inflow histories result in the observed correlation between metallicity and gas mass, with \emph{metallicity depending on the rate of change of gas mass} for galaxies. Results from the literature additionally suggest that galaxy inflow metallicities are relevant, with gas-poor galaxies expected to receive a greater proportion of their inflow from enriched recycled gas, increasing their metallicity further. 

In our proposed view, more extended star-forming galaxies exist within more massive dark halos, resulting in them accreting significant quantities of metal-poor gas; this results in lower metallicities and more massive gas reservoirs. By contrast, compact galaxies exist within less massive dark halos, resulting in a steadily decreasing supply of metal-poor gas,  which in turn results in inflows becoming dominated by enriched recycled gas; this results in less massive gas reservoirs and higher metallicities, with the metallicity sensitive both to the long-term inflow history and the current inflow metallicity.

Our results provide critical tests for models and simulations, with particular examples being dark halo masses across stellar mass--size space and the close connection between metallicity and gas mass within galaxies' optical extents. Our results also open up a number of avenues for further exploration. Resolved HI maps will allow for much more precise investigations of HI across the mass-size plane, and this will for instance be possible with WEAVE-Apertif. It will also be important to investigate how HI mass connects to metallicity at higher redshifts, which will for instance be possible with SKA. 

\section*{Acknowledgments}

We thank the anonymous referee for their thoughtful and constructive review, which significantly improved this paper.

NFB and VW acknowledge support from Science and Technologies Facilities Council (STFC) grant ST/Y00275X/1. VW acknowledges support from Leverhulme Research Fellowship (RF-2024-589/4). KW acknowledges support from the STFC through grant ST/X001075/1. NVA acknowledges support of Conselho Nacional de Desenvolvimento Cient\'{\i}fico e Tecnol\'{o}gico (CNPq). This study was financed in part by the Coordena\c{c}\~{a}o de Aperfei\c{c}oamento de Pessoal de N\'{\i}vel Superior - Brasil (CAPES) -- Finance Code 001. NFB thanks Ricardo Schiavon for introducing him to VICE along with other helpful discussions.

SDSS-IV is managed by the Astrophysical Research Consortium for the Participating Institutions of the SDSS Collaboration including the Brazilian Participation Group, the Carnegie Institution for Science, Carnegie Mellon University, the Chilean Participation Group, the French Participation Group, Harvard-Smithsonian Center for Astrophysics, Instituto de Astrof\'isica de Canarias, The Johns Hopkins University, Kavli Institute for the Physics and Mathematics of the Universe (IPMU) / University of Tokyo, Lawrence Berkeley National Laboratory, Leibniz Institut f\"ur Astrophysik Potsdam (AIP),  Max-Planck-Institut f\"ur Astronomie (MPIA Heidelberg), Max-Planck-Institut f\"ur Astrophysik (MPA Garching), Max-Planck-Institut f\"ur Extraterrestrische Physik (MPE), National Astronomical Observatories of China, New Mexico State University, New York University, University of Notre Dame, Observat\'ario Nacional / MCTI, The Ohio State University, Pennsylvania State University, Shanghai Astronomical Observatory, United Kingdom Participation Group, Universidad Nacional Aut\'onoma de M\'exico, University of Arizona, University of Colorado Boulder, University of Oxford, University of Portsmouth, University of Utah, University of Virginia, University of Washington, University of Wisconsin, Vanderbilt University, and Yale University.

\section*{Data Availability}

All data used here are publicly available. MaNGA data and analysis products can be accessed through the \textsc{marvin} interface online or else with \textsc{python}\footnote{\url{https://www.sdss4.org/dr17/manga/marvin/}}. MaNGA data and products can also be downloaded from the SDSS Science Archive server\footnote{\url{https://data.sdss.org/sas/}}. All value added catalogs used in this article (HI-MaNGA, \textsc{pypipe3d}, MaNGA DynPop) are also publically available. A \rev{git repository} containing this article's analysis and plotting code is available online at \url{https://github.com/NickyBfudd/boardman2026_source}.

\bibliography{bibliography}

\appendix

\section{Alternative metallicity calibrators}\label{altcallib}

Here, we briefly explore the behaviour of two other metallicity calibrators: the \citet{curti2020} RS32 calibrator and the \citet{marino2013} O3N2 calibrator. The RS32 calibrator has the advantage of being less sensitive to one's choice of dust correction and also allows one to investigate N/O without correlated uncertainties \citep{luo2021,boardman2024a}, though a downside is that it implicitly assumes a fixed S/O ratio which may in practice be inaccurate \citep{perezdiaz2024}. O3N2 is almost completely insensitive to dust attenuation meanwhile, though in practice it is more indicative of N/O than of metallicity \citep{florido2022}.

\begin{figure}
\begin{center}
	\includegraphics[trim = 0cm 0.3cm 0cm 0.1cm,scale=0.8]{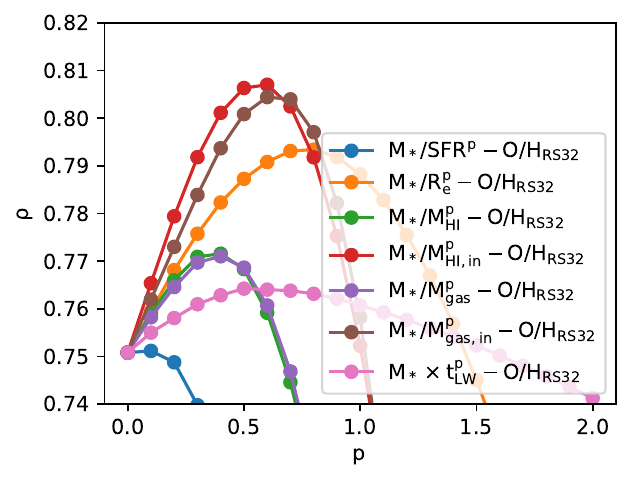}
    \includegraphics[trim = 0cm 1cm 0cm 0cm,scale=0.8]{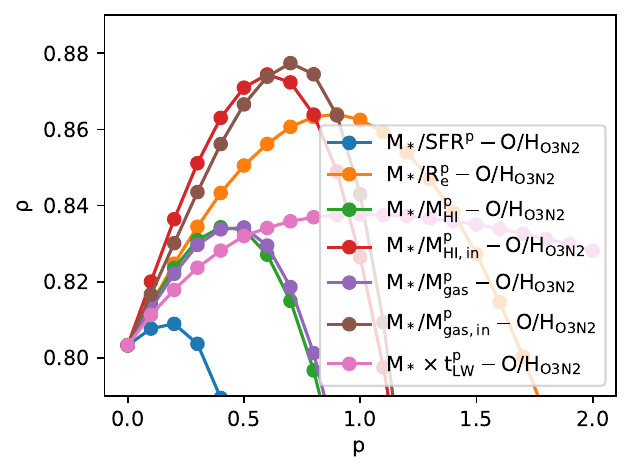}
	\caption{As in Figure \ref{scorrs}, but for metallicities derived with the \citet{curti2020} RS32 calibrator (top) and with the \citet{marino2013} O3N2 calibrator (bottom). We obtain similar results as for R23, indicating that calibrator choice does not drive our findings.}
	\label{scorrs_alt}
	\end{center}
\end{figure}

We present in Figure \ref{scorrs_alt} a series of Spearman correlation coefficients in the same format as in Figure \ref{scorrs}, \rev{with the same secondary parameters (SFR, $\mathrm{R_e}$, $\mathrm{M_{HI}}$, $\mathrm{M_{HI,in}}$, $\mathrm{M_{gas}}$, $\mathrm{M_{gas,in}}$, $\mathrm{t_{LW}}$)} considered as before. These plots were constructed using the same 1542 galaxies as in the main paper text. For O3N2, we find entire equivalent results to R23: we find $\mathrm{M_{gas,in}}$ to be the most constraining parameter after $\mathrm{M_*}$, followed closely by $\mathrm{M_{HI,in}}$ and then followed $\mathrm{R_e}$. For RS32, we instead find $\mathrm{M_{HI,in}}$ to yield the strongest correlation, though this is then followed closely by $\mathrm{M_{gas,in}}$. \rev{$\mathrm{t_{LW}}$ also produces weaker correlations in this case, relative to all tested parameters aside from SFR}. The differences in performance between $\mathrm{M_{HI,in}}$ and $\mathrm{M_{gas,in}}$ are marginal regardless of metallicity calibrator, so our overall discussion and interpretation is not affected by which of the two performs the strongest. The correlation \rev{coefficients} are generally somewhat higher for O3N2 than for RS32 or R23, which can be understood as being due to O3N2's connection to the N/O ratio (Figure \ref{scorrs_no}, top panel).

\section{Plots without smoothing}\label{nosmooth}

Within our article, we repeatedly used 2D smoothing to emphasize two-dimensional trends across various parameter spaces. Here, we present equivalent plots without smoothing. 

\begin{figure}
\begin{center}
    \includegraphics[trim = 0.3cm 0.3cm 0cm 0cm,scale=0.75]{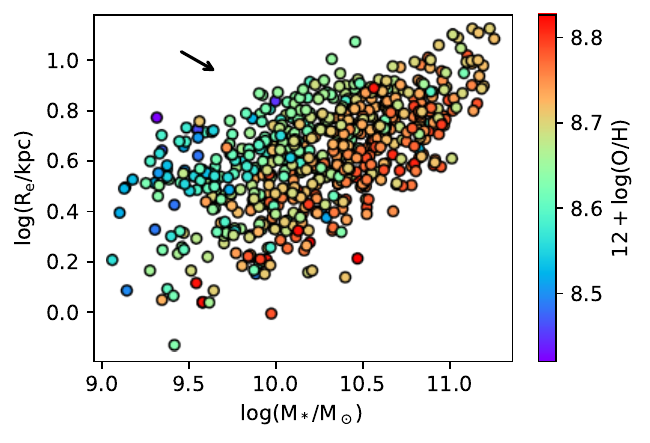} 
	\includegraphics[trim = 0cm 0.5cm 0cm 0cm,scale=0.75]{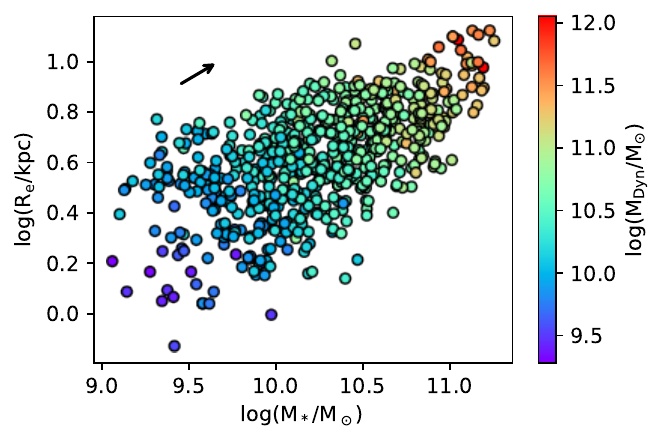} 
	\caption{Stellar mass ($\mathrm{M_*}$ vs half-light radius for the $\mathrm{M_{Dyn}}$ subsample, coloured by gas metallicity (top) and by dynamical mass within $\mathrm{1 R_e}$ (bottom). This is a smoothing-free version of Figure \ref{mre_dyn}.}
	\label{mre_dyn_nosmooth}
	\end{center}
\end{figure}

\begin{figure}
\begin{center}
	\includegraphics[trim = 0.3cm 0cm 0cm 0.2cm,scale=0.75]{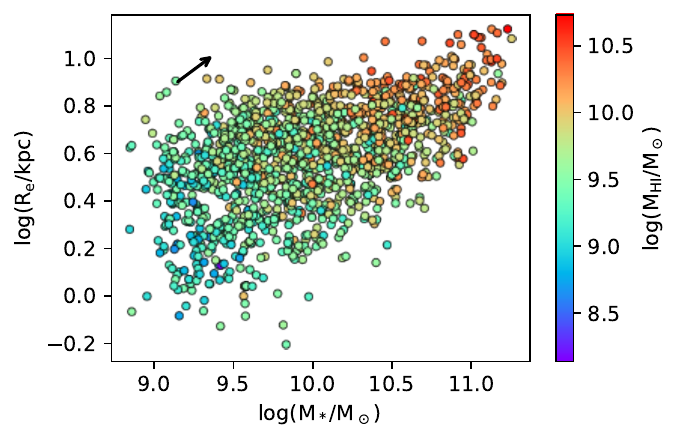} 
    \includegraphics[trim = 0.3cm 0.7cm 0cm 0.2cm,scale=0.75]{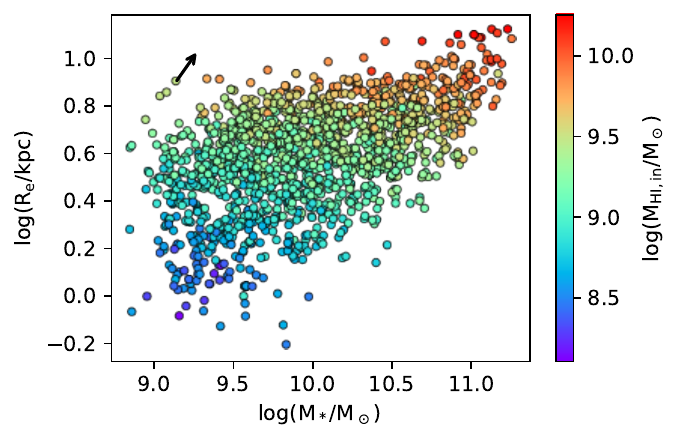} 
	\caption{Top: global HI gas mass ($\mathrm{M_{HI}}$) plotted as a combined function of stellar mass ($\mathrm{M_*}$) and half-light radius ($\mathrm{R_e}$. The arrow shows the direction of maximum $\mathrm{M_{HI}}$ increase, computed using partial correlation coefficients. Bottom: as above, but for the HI mass within galaxies' 90\% light radii ($\mathrm{M_{HI,in}}$). These are smoothing-free versions of the top two panels in Figure \ref{mhismooth}.}
	\label{mhinosmooth}
	\end{center}
\end{figure}

\begin{figure}
\begin{center}
	\includegraphics[trim = 0.2cm 0.1cm 0cm 0.2cm,scale=0.75]{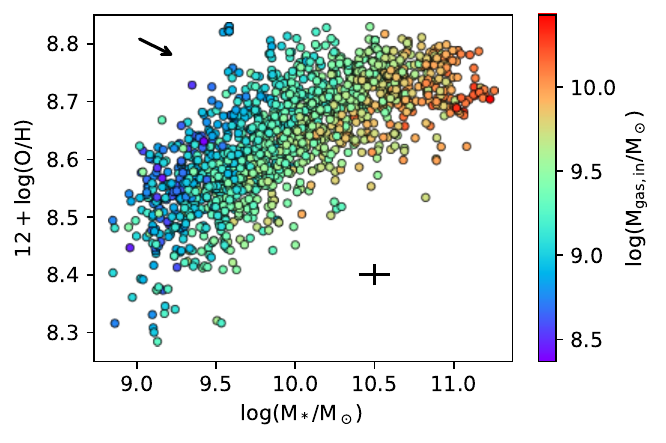} 
    \includegraphics[trim = 0.3cm 0.1cm 0cm 0.2cm,scale=0.75]{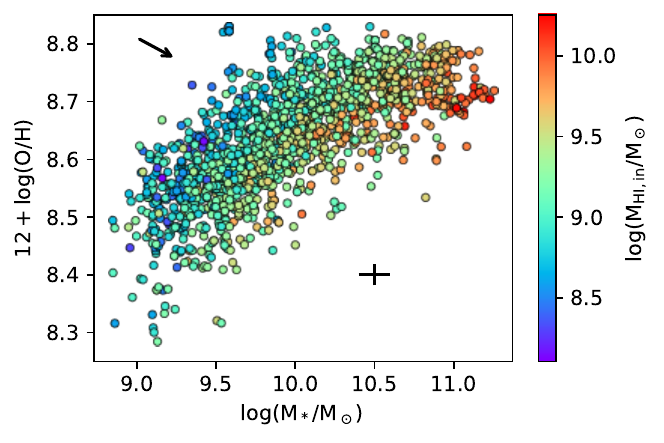} 
    \includegraphics[trim = 0.2cm 0.7cm 0cm 0.2cm,scale=0.75]{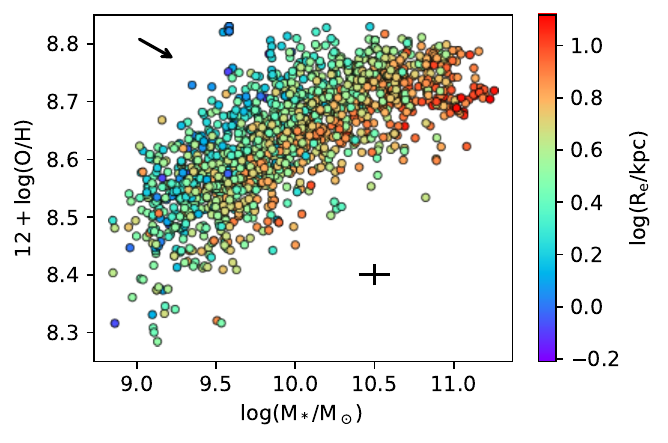} 
	\caption{Mass--metallicity relation colored by $\mathrm{M_{gas,in}}$ (top), $\mathrm{M_{HI,in}}$ (middle) or $\mathrm{R_e}$ (bottom). The error bars show the median errors in stellar mass and metallicity, and the arrows show directions of maximum increase. This is a smoothing-free version of Figure \ref{msmgoh}.}
	\label{msmgohnosmooth}
	\end{center}
\end{figure}

\begin{figure}
\begin{center}
	\includegraphics[trim = 0cm 1cm 0cm 0cm,scale=0.75]{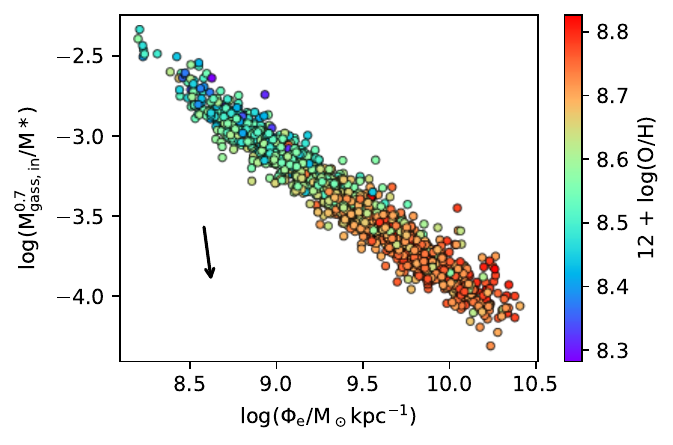} 
	\caption{$\mathrm{M_{gas,in}^{0.7}/M_*}$ vs $\Phi_e$, colored by gas metallicity. We also show the direction of maximum increase, computed from partial correlation coefficients. This is a smoothing-free version of Figure \ref{inmgas_phi_pcc}}
	\label{inmgas_phi_pcc_nosmooth}
	\end{center}
\end{figure}

\begin{figure}
\begin{center}
	\includegraphics[trim = 0cm 0cm 0cm 0cm,scale=0.75]{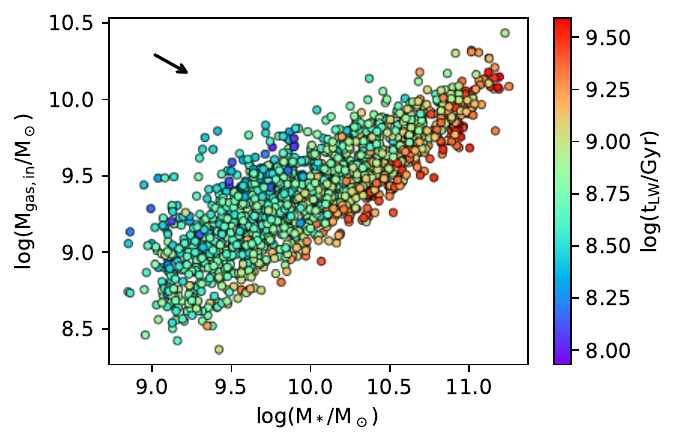} 
	\caption{Inner gas mass vs stellar mass, colored by light-weighted stellar ages at 1~$\mathrm{R_e}$. We also show the direction of maximum increase, computed from partial correlation coefficients. This is a smoothing-free version of Figure \ref{MsMg_age}.}
	\label{MsMg_age_nosmooth}
	\end{center}
\end{figure}

We show in Figure \ref{mre_dyn_nosmooth} our $\mathrm{M_{dyn}}$ subsample galaxies' metallicity and $\mathrm{M_{Dyn}}$ as combined functions of $\mathrm{M_*}$ and $\mathrm{R_e}$, as in Figure \ref{mre_dyn}. In Figure \ref{mhinosmooth}, we present $\mathrm{M_{HI}}$ and $\mathrm{M_{HI,in}}$ as combined functions of $\mathrm{M_*}$ and $\mathrm{R_e}$, as in the top two plots of Figure \ref{mhismooth}. In Figure \ref{msmgohnosmooth}, we present the mass-metallicity relation coloured by $\mathrm{M_{gas,in}}$, $\mathrm{M_{HI,in}}$ and $\mathrm{R_e}$, as in Figure \ref{msmgoh}. We show in Figure \ref{inmgas_phi_pcc_nosmooth} the gas metallicity as a combined function of $\mathrm{M_{gas,in}^{0.7}/M_*}$ $\Phi_e$, as in Figure \ref{inmgas_phi_pcc}. Finally, we show in Figure \ref{MsMg_age_nosmooth} galaxies' light-weighted stellar ages at 1~$\mathrm{R_e}$ as a combined function of $\mathrm{M_*}$ and $\mathrm{M_{gas,in}}$, as in Figure \ref{MsMg_age}

\section{Random forests and $\mathrm{\Phi_e}$}\label{rfreduced}

Over the course of our article, we established a strong connection between metallicity and $\mathrm{M_{gas,in}}$ at any given stellar mass. Using Spearman correlation coefficients, we then found the parameter $\mathrm{M_{gas,in}^{0.7}/M_*}$ to yield the tightest correlation with metallicity out of all tested parameters. A random forest analysis confirmed this view, with $\mathrm{M_{gas,in}^{0.7}/M_*}$ obtaining by far the highest importance for determining metallicity. We found $\Phi_e$ to carry relatively little importance in the random forest, despite the known close correlation between $\Phi_e$ and metallicity \citep[e.g.][]{deugenio2018,sm2024}; we argued this to be due to the close correlation between $\Phi_e$ and $\mathrm{M_{gas,in}^{0.7}/M_*}$, with metallicity correlating more closely with the latter parameter.

\begin{figure*}
\begin{center}
	\includegraphics[trim = 0cm 0.5cm 0cm 0cm,scale=0.85]{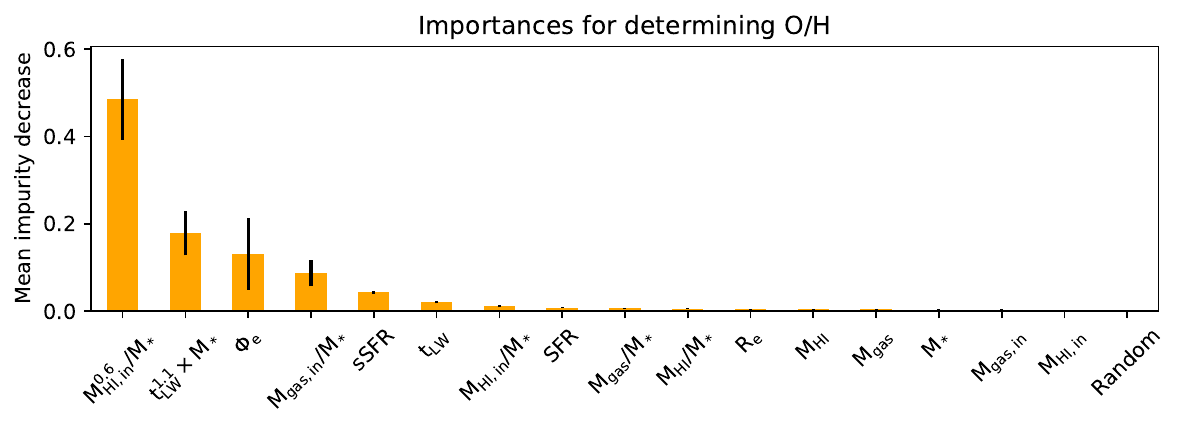} 
	\caption{Feature importances for determining gas metallicity from a second random forest analysis in which $\mathrm{M_{gas,in}^{0.7}/M_*}$ was excluded as a feature. We report importances as the means across 50 RF realisations, with errors reported as standard deviations. We now find $\mathrm{M_{HI,in}^{0.6}/M_*}$  to be the most important parameter, with other parameters continuing to attain comparitively little importance.}
	\label{randomforest_lessparams}
	\end{center}
\end{figure*}

In Figure \ref{randomforest_lessparams}, we show feature importances from an additional random forest run where $\mathrm{M_{gas,in}^{0.7}/M_*}$ was excluded as a feature; this was done with the same RF settings as before, for R23-based metallicities. The RF performance is very similar to the previous run, with average root-mean-square errors of \rev{0.041 and 0.049} on the training and test sets. We find that $\mathrm{M_{HI,in}^{0.6}/M_*}$ carries the most importance in this case, \rev{with $\mathrm{t_{LW}^{1.1} \times M_*}$ found to be the second-most important parameter like before.} 

We now also exclude $\mathrm{M_{HI,in}^{0.6}/M_*}$ and perform an additional RF run, with the results shown in Figure \ref{randomforest_{LW}venlessparams}. We continue to achieve similar RF performnace, with root-mean-square errors of {0.041 and 0.049} on the training and test sets. We now find that $\Phi_e$ carries the highest importance, in agreement with the results of \citet{sm2024}. This does \textit{not}, however, imply that $\Phi_e$ connects fundamentally to a galaxy's metallicity. Rather, this can be understood as being due to $\Phi_e$ closely reflecting a galaxy's position in $\mathrm{M_*}$--$\mathrm{M_{gas,in}}$ parameter space (see Figure 8 in the main paper text), leading the random forest to highly weight $\Phi_e$ at the expense of parameters that directly relate to the gas reservoir. \rev{That $\mathrm{t_{LW}^{1.1} \times M_*}$ consistently achieves the second-highest importance suggests that this parameter holds complementary information, further suggesting a connection between galaxy metallicities and galaxies' long-term formation histories.}

\rev{In summary: when we don't include the most informative feature ($\mathrm{M_{gas,in}^{0.7}/M_*}$, then $\mathrm{M_{HI,in}^{0.6}/M_*}$) from the random forest, the importance of a closely correlated feature ($\Phi_e$) is boosted while the importance of less correlated features (e.g. sSFR) is affected far more mildly. This emphasizes the need to include the most informative features in a RF analysis to avoid misleading outcomes.} 

\begin{figure*}
\begin{center}
	\includegraphics[trim = 0cm 0.5cm 0cm 0cm,scale=0.85]{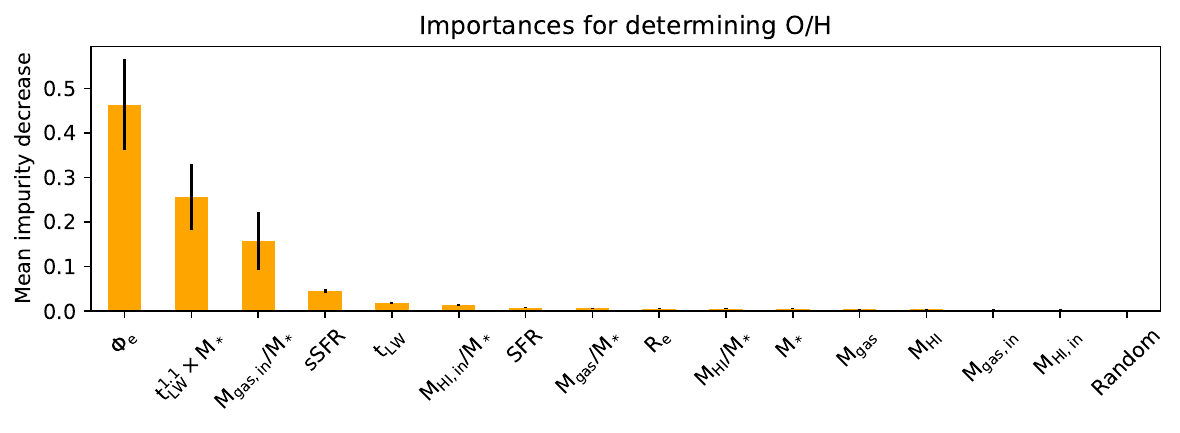} 
	\caption{Feature importances for determining gas metallicity from third random forest analysis in which $\mathrm{M_{gas,in}^{0.7}/M_*}$ and $\mathrm{M_{HI,in}^{0.6}/M_*}$ were excluded as a feature. We now find $\Phi_e$ to be the most important parameter, consistent with previous work; we argue this to be due to $\Phi_e$ serving as a close proxy for a galaxy's position in $\mathrm{M_{gas,in}}$--$\mathrm{M_*}$ space, leading to it being up-weighted over parameters that directly relate to gas mass..}
	\label{randomforest_{LW}venlessparams}
	\end{center}
\end{figure*}

\label{lastpage}
\end{document}